\documentclass[aps,showpacs,floatfix]{revtex4}

\usepackage{amsmath}
\usepackage{amssymb}
\usepackage{graphicx}

\begin{document}

\title{Can dark matter be a Bose-Einstein condensate?}

\author{C. G. B\"ohmer}
\email{christian.boehmer@port.ac.uk}
\affiliation{Institute of Cosmology \& Gravitation,
             University of Portsmouth, Portsmouth PO1 2EG, UK}

\author{T. Harko}
\email{harko@hkucc.hku.hk}
\affiliation{Department of Physics and Center for Theoretical and
             Computational Physics, The University of Hong Kong,
             Pok Fu Lam Road, Hong Kong}

\date{\today}

\begin{abstract}
We consider the possibility that the dark matter, which is
required to explain the dynamics of the neutral hydrogen clouds at
large distances from the galactic center, could be in the form of
a Bose-Einstein condensate. To study the condensate we use the
non-relativistic Gross-Pitaevskii equation. By introducing the
Madelung representation of the wave function, we formulate the
dynamics of the system in terms of the continuity equation and of
the hydrodynamic Euler equations. Hence dark matter can be
described as a non-relativistic, Newtonian Bose-Einstein
gravitational condensate gas, whose density and pressure are
related by a barotropic equation of state. In the case of a
condensate with quartic non-linearity, the equation of state is
polytropic with index $n=1$. In the framework of the Thomas-Fermi
approximation the structure of the Newtonian gravitational
condensate is described by the Lane-Emden equation, which can be
exactly solved. General relativistic configurations with quartic
non-linearity are studied, by numerically integrating the
structure equations. The basic parameters (mass and radius) of the
Bose-Einstein condensate dark matter halos sensitively depend on
the mass of the condensed particle and of the scattering length.
To test the validity of the model we fit the Newtonian tangential
velocity equation of the model with a sample of rotation curves of
low surface brightness and dwarf galaxies, respectively. We find a
very good agreement between the theoretical rotation curves and
the observational data for the low surface brightness galaxies.
The deflection of photons passing through the dark matter halos is
also analyzed, and the bending angle of light is computed. The
bending angle obtained for the Bose-Einstein condensate is larger
than that predicted by standard general relativistic and dark
matter models. The angular radii of the Einstein rings are
obtained in the small angles approximation. Therefore the study of
the light deflection by galaxies and the gravitational lensing
could discriminate between the Bose-Einstein condensate dark
matter model and other dark matter models.
\end{abstract}

\pacs{04.50.+h, 04.20.Jb, 04.20.Cv, 95.35.+d}

\maketitle

\section{Introduction}

The existence of the dark matter in the Universe is a well
established observational fact. The rotation curves of spiral
galaxies \cite{Bi87} are one of the best evidences showing a
modified gravitational particle dynamic at galactic level. In
these galaxies neutral hydrogen clouds are observed at large
distances from the center, much beyond the extent of the luminous
matter. Assuming a non-relativistic Doppler effect and emission
from stable circular orbits in a Newtonian gravitational field,
the frequency shifts in the $21$ cm line hydrogen emission lines
allows the measurement of the velocity of the clouds. Since the
clouds move in circular orbits with velocity $v_{tg}(r)$, the
orbits are maintained by the balance between the centrifugal
acceleration $v_{tg}^2/r$ and the gravitational attraction force
$GM(r)/r^2$ of the total mass $M(r)$ contained within the orbit.
This allows the expression of the mass profile of the galaxy in
the form $M(r)=rv_{tg}^2/G$.

Observations show that the rotational velocities increase near the
center of the galaxy and then remain nearly constant at a value of
$v_{tg\infty }\sim 200$ km/s \cite{Bi87}. This leads to a mass
profile $M(r)=rv_{tg\infty }^2/G$. Consequently, the mass within a
distance $r$ from the center of the galaxy increases linearly with
$r$, even at large distances where very little luminous matter can
be detected.

Several theoretical models, based on a modification of Newton's
law or of general relativity, have been proposed to explain the
behavior of the galactic rotation curves. A modified gravitational
potential of the form $\phi =-GM\left[ 1+\alpha \exp \left(
-r/r_{0}\right) \right] /\left( 1+\alpha \right) r$, with $\alpha
=-0.9$ and $r_{0}\approx 30$ kpc can explain flat rotational
curves for most of the galaxies \cite{Sa84}.

In an other model, called MOND, and proposed by Milgrom \cite{Mi},
the Poisson equation for the gravitational potential $\nabla
^{2}\phi =4\pi G\rho $ is replaced by an equation of the form
$\nabla \left[ \mu \left( x\right) \left( \left| \nabla \phi
\right| /a_{0}\right) \right] =4\pi G\rho $, where $a_{0}$ is a
fixed constant and $\mu \left( x\right) $ a function satisfying
the conditions $\mu \left( x\right) =x$ for $x \ll 1$ and $\mu \left(
x\right) =1$ for $x \gg 1$. The force law, giving the acceleration
$a$ of a test particle  becomes $a=a_{N}$ for $a_{N} \gg a_{0}$ and $a=\sqrt{%
a_{N}a_{0}}$ for $a_{N} \ll a_{0},$where $a_{N}$ is the usual
Newtonian acceleration. The rotation curves of the galaxies are
predicted to be flat, and they can be calculated once the
distribution of the baryonic matter is known. Alternative
theoretical models to explain the galactic rotation curves have
been elaborated recently by Mannheim \cite{Ma93}, Moffat and
Sokolov \cite{Mo96} and Roberts \cite{Ro04}. The idea that dark
matter is a result of the bulk effects in brane world cosmological
models was considered in \cite{Ha}.

A general analysis of the possibility of an alternative gravity
theory explaining the dynamics of galactic systems without dark
matter was performed by Zhytnikov and Nester \cite{Ne94}. From
very general assumptions about the structure of a relativistic
gravity theory (the theory is metric, invariant under general
coordinates transformation, has a good linear approximation, it
does not possess any unusual gauge freedom and it is not a higher
derivative gravity), a general expression for the metric to order
$(v/c)^2$ has been derived. This allows to compare the predictions
of the theory with various experimental data: the Newtonian limit,
light deflection and retardation, rotation of galaxies and
gravitational lensing. The general conclusion of this study is
that the possibility for any gravity theory to explain the
behavior of galaxies without dark matter is rather improbable.

Hence one of most promising ways to explain the galactic rotation
curves is by postulating the existence of some dark (invisible)
matter, distributed in a spherical halo around the galaxies. This
behavior of the galactic rotation curves is explained by
postulating the existence of some dark (invisible) matter,
distributed in a spherical halo around the galaxies. The dark
matter is assumed to be a cold, pressure-less medium. There are
many possible candidates for dark matter, the most popular ones
being the weekly interacting massive particles (WIMP) (for a
recent review of the particle physics aspects of dark matter see
\cite{OvWe04}). Their interaction cross section with normal
baryonic matter, while extremely small, are expected to be
non-zero and we may expect to detect them directly. It has also
been suggested that the dark matter in the Universe might be
composed of superheavy particles, with mass $\geq 10^{10}$ GeV.
But observational results show the dark matter can be composed of
superheavy particles only if these interact weakly with normal
matter or if their mass is above $10^{15}$ GeV \cite{AlBa03}.
Scalar fields or other long range coherent fields coupled to
gravity have also intensively been used to model galactic dark
matter \cite{Ma03}.

At very low temperatures, particles in a dilute Bose gas can
occupy the same quantum ground state, forming a Bose-Einstein
(BEC) condensate, which appears as a sharp peak over a broader
distribution in both coordinates and momentum space. The
possibility to obtain quantum degenerate gases by a combination of
laser and evaporative cooling techniques has opened several new
lines of research, at the border of atomic, statistical and
condensed matter physics (for recent reviews see \cite{Da99,rev}).

An ideal system for the experimental observation of the BEC
condensation is a dilute atomic Bose gas confined in a trap and
cooled to very low temperatures. BEC were first observed in 1995
in dilute alkali gases such as vapors of rubidium and sodium. In
these experiments, atoms were confined in magnetic traps,
evaporatively cooled down to a fraction of a microkelvin, left to
expand by switching off the magnetic trap, and subsequently imaged
with optical methods. A sharp peak in the velocity distribution
was observed below a critical temperature, indicating that
condensation has occurred, with the alkali atoms condensed in the
same ground state. Under the typical confining conditions of
experimental settings, BEC's are inhomogeneous, and hence
condensates arise as a narrow peak not only in the momentum space
but also in the coordinate space \cite{exp}.

If considering only two-body, mean field interactions, a dilute
Bose-Einstein gas near zero temperature can be modeled using a
cubic non-linear Schr\"odinger equation with an external
potential, which is known as the Gross-Pitaevskii equation
\citep{Da99}.

The idea that the dark matter is composed of ultra-light scalar
particles ($m \approx 10^{-22}$ eV) in a (cold) Bose-Einstein
condensate state, was proposed initially in \cite{sin}, and
further developed in \cite{Hu00}. The wave properties of the dark
matter stabilize gravitational collapse, providing halo cores and
sharply suppressing small-scale linear power. The speed of BEC's
in atomic vapors and in galactic dark matter was studied in
\cite{Kl02}.  A cosmological model in which the boson dark matter
gradually condensates into dark energy was considered in
\cite{Ni04}. Negative pressure associated with the condensate
yields the accelerated expansion of the Universe and the rapid
collapse of the smallest scale fluctuations into many black holes,
which become the seeds of the first galaxies.

Scalar mediated interactions among baryons embedded in a
Bose-Einstein condensate, composed of the mediating particles,
which extend well beyond the Compton wavelength, were studied in
\cite{Fe04}. If the dark matter of the Universe is composed of
such a condensate, the imprints of an interaction between the
baryonic and the dark matter could be manifest as anomalies in the
peak structure of the Cosmic Microwave Background. In a medium
composed of scalar particles with non-zero mass, the range of
Van der Waals-type scalar mediated interactions among nucleons
becomes infinite when the medium makes a transition to a
Bose-Einstein condensed phase. In an astrophysical context this
phenomenon  was explored in \cite{Gr06} and the effect of a scalar
dark matter background on the equilibrium of degenerate stars was
studied.

A relativistic version of the Gross-Pitaevskii equation was
proposed in \cite{Fu05} and the cosmological implications of a
steady slow BEC process were considered. It is interesting to note
that the resulting equation of state for the condensate is that of
the Whittaker solution \cite{Wh68}, which has special importance
in classical general relativity. Namely, the solution is the
static non-rotating limit of the Wahlquist solution, which is
probably the most important exact rotating perfect fluid
spacetime.

Dolgov and Smirnov \cite{DoSm05} assumed that the Pauli exclusion
principle is violated for neutrinos, and consequently, neutrinos
obey the Bose-Einstein statistics. Neutrinos may form cosmological
Bose condensates, which accounts for all (or a part of) the dark
matter in the universe. ``Wrong'' statistics of neutrinos could
modify big bang nucleosynthesis, leading to the effective number
of neutrino species smaller than three. The Pauli principle
violation for neutrinos can be tested in the two-neutrino double
beta decay. In order to effectively modify Kepler's law without
changing standard Newtonian gravity, the galactic dark matter was
described by a scalar field  in \cite{MiFuSc06}. For cold scalar
fields, this model corresponds to a gravitationally confined
Boson-Einstein condensate, but of galactic dimensions. A light,
neutral vector particle associated with a vector field $\phi^\mu$,
and which appears in a modified theory of gravity, may form a cold
fluid of Bose-Einstein condensates before the cosmological
recombination with zero pressure and zero shear viscosity
\cite{Mo06}.  Vortices in axion condensates on the galactic scale
have been studied in \cite{Sz07}. Such vortices can occur as a
result of global rotation of the early universe. Various
cosmological implications of axion condensation have been
investigated in \cite{Khl}.

The dynamical equations describing the evolution of a
self-gravitating fluid of cold dark matter can be written in the
form of a Schrodinger equation coupled to a Poisson equation,
describing Newtonian gravity. It has been shown that, in the
quasi-linear regime, the Schrodinger equation can be reduced to
the exactly solvable free-particle Schrodinger equation. This
approach can be used to study gravitational instabilities or
structure formation \cite{coles}. Unified models for dark matter
and dark energy via a single fluid and models discussing the
properties of the dark energy and cosmological horizons have been
proposed and discussed in \cite{Arb}.

It is the purpose of the present paper to systematically
investigate the possibility that the dark matter, which is
required to explain the dynamics of the neutral hydrogen clouds at
large distances from the galactic center, could be in the form of
a Bose-Einstein condensate. To study the dark matter condensate we
use the non-relativistic Gross-Pitaevskii equation. By introducing
the Madelung representation of the wave function, we formulate the
dynamics of the (quantum) system in terms of the continuity
equation and of the hydrodynamic Euler equations. Hence dark
matter can be described as a non-relativistic, Newtonian
Bose-Einstein gravitational condensate gas, whose density and
pressure are related by a barotropic equation of state. In the
case of a condensate with quartic non-linearity, the equation of
state is polytropic with index $n=1$.

In the framework of the Thomas-Fermi approximation, the structure
of the Newtonian gravitational condensate is described by the
Lane-Emden equation, which can be exactly solved. General
relativistic configurations with quartic non-linearity are
studied, by numerically integrating the structure equations. The
basic parameters (mass and radius) of the Bose-Einstein condensate
dark matter halos sensitively depend on the mass $m$ of the
condensed particle, of the scattering length $a$ and of the
central density of the dark matter. The values of these parameters
can be constrained by using the known radii and masses of the
galactic dark matter halos.

To test the viability of the model and in order to apply it to
realistic systems, we compute the rotation curve of the
Bose-Einstein condensate for a sample of 12 galaxies that includes
low surface brightness and dwarf galaxies with measured rotation
curves extending in the dark matter dominated region.  A best fit
to the rotation curves of galaxies is obtained in terms of a
parametric core baryonic mass distribution, which is superposed
with the Bose-Einstein condensate mass distribution. It turns out
that the Newtonian potential of the core is asymptotically
decreasing, but the corrected rotation curve is much higher than
the Newtonian one, thus offering the possibility to fit the
rotation curves. For low surface brightness galaxies we find a
very good agreement between the theoretical rotation curve with a
normal baryonic core and the observational data.

The deflection of photons passing through the dark matter halos is
also analyzed, and the bending angle of light is computed. The
bending angle obtained for the Bose-Einstein condensate is larger
than that predicted by standard general relativistic and dark
matter models. The angular radii of the Einstein rings are
obtained in the small angles approximation. Therefore the study of
the light deflection by galaxies and the gravitational lensing
provides a powerful observational method to discriminate between
the Bose-Einstein condensate dark matter model and other dark
matter models.

The present paper is organized as follows. The basic equations
describing the gravitationally bounded Bose-Einstein condensate
are derived in Section II. The case of the condensate with quartic
non-linearity is considered in Section III. Newtonian dark matter
condensate models are analyzed in Section IV. General relativistic
condensate models are studied numerically in Section V. We compare
the predictions of our model with the observed galactic rotation
curves in Section VI. The bending of light by dark matter
condensate halos is considered in Section VII. We discuss and
conclude our results in Section VIII.

\section{The Gross-Pitaevskii equation for gravitationally bounded Bose-Einstein condensates}

In a quantum system of $N$ interacting condensed bosons most of
the bosons lie in the same single-particle quantum state. For a
system consisting of a large number of particles, the calculation
of the ground state of the system with the direct use of the
Hamiltonian is impracticable, due to the high computational cost.
However, the use of some approximate methods can lead to a
significant simplification of the formalism. One such approach is
the mean field description of the condensate, which is based on
the idea of separating out the condensate contribution to the
bosonic field operator. We also assume that in a medium composed
of scalar particles with non-zero mass, the range of Van der
Waals-type scalar mediated interactions among nucleons becomes
infinite, when the medium makes a transition to a Bose-Einstein
condensed phase.

The many-body Hamiltonian describing the interacting bosons
confined by an external potential $V_{ext}$ is given, in the
second quantization, by
\begin{align}
\hat{H}=\int d\vec{r}\hat{\Psi}^{+}\left( \vec{r}\right) \left[ -\frac{%
\hbar ^{2}}{2m}\nabla ^{2}+V_{rot}\left( \vec{r}\right) +V_{ext}\left( \vec{r%
}\right) \right] \hat{\Psi}\left( \vec{r}\right) +
\frac{1}{2}\int d\vec{r}d\vec{r}^{\prime }\hat{\Psi}^{+}\left( \vec{r}%
\right) \hat{\Psi}^{+}\left( \vec{r}^{\prime }\right) V\left( \vec{r}-\vec{r}%
^{\prime }\right) \hat{\Psi}\left( \vec{r}\right) \hat{\Psi}\left( \vec{r}%
^{\prime }\right) ,  \label{ham}
\end{align}
where $\hat{\Psi}\left( \vec{r}\right) $ and $\hat{\Psi}^{+}\left( \vec{r}%
\right) $ are the boson field operators that annihilate and create
a
particle at the position $\vec{r}$, respectively, and $V\left( \vec{r}-\vec{r%
}^{\prime }\right) $ is the two-body interatomic potential \cite{Da99}. $%
V_{rot}\left( \vec{r}\right) $ is the potential associated to the
rotation of the condensate, and is given by
\begin{align}
V_{rot}\left( \vec{r}\right) =f_{rot}\left( t\right) \frac{m\omega ^{2}}{2}%
\vec{r}^{2},
\end{align}
where $\omega $ is the angular velocity of the condensate and
$f_{rot}\left( t\right) $ a function which takes into account the
possible time variation of the rotation potential. For a system
consisting of a large number of particles, the calculation of the
ground state of the system with the direct use of Eq. (\ref{ham})
is impracticable, due to the high computational cost.

Therefore, the use of some approximate methods can lead to a
significant simplification of the formalism. One such approach is
the mean field description of the condensate, which is based on
the idea of separating out the condensate contribution to the
bosonic field operator. For a uniform gas in a volume $V$, BEC
occurs in the single particle state $\Psi _{0}=1\sqrt{V} $, having
zero momentum. The field operator can be decomposed then in the
form $\hat{\Psi}\left( \vec{r}\right) =\sqrt{N/V}+\hat{\Psi}^{\prime }\left(%
\vec{r}\right) $. By treating the operator $\hat{\Psi}^{\prime }\left( \vec{r%
}\right) $ as a small perturbation, one can develop the first
order theory for the excitations of the interacting Bose gases
\cite{Da99}.

In the general case of a non-uniform and time-dependent
configuration the field operator in the Heisenberg representation
is given by
\begin{align}
\hat{\Psi}\left( \vec{r},t\right) =\psi \left( \vec{r},t\right) +\hat{\Psi}%
^{\prime }\left( \vec{r},t\right) ,
\end{align}
where $\psi \left( \vec{r},t\right) $, also called the condensate
wave function, is the expectation value of the field operator, $\psi \left( \vec{r%
},t\right) =\left\langle \hat{\Psi}\left( \vec{r},t\right)
\right\rangle $. It is a classical field and its absolute value
fixes the number density of the condensate through $\rho \left(
\vec{r},t\right) =\left| \psi \left( \vec{r},t\right) \right|
^{2}$. The normalization condition is $N=\int \rho \left(
\vec{r},t\right) d^{3}\vec{r}$, where $N$ is the total number of
particles in the condensate.

The equation of motion for the condensate wave function is given
by the
Heisenberg equation corresponding to the many-body Hamiltonian given by Eq. (%
\ref{ham}),
\begin{align}
i\hbar \frac{\partial }{\partial t}\hat{\Psi}\left( \vec{r},t\right) =%
\left[ \hat{\Psi},\hat{H}\right] =
\left[ -\frac{\hbar ^{2}}{2m}\nabla ^{2}+V_{rot}\left(
\vec{r}\right) +V_{ext}\left( \vec{r}\right) +\int
d\vec{r}^{\prime }\hat{\Psi}^{+}\left(
\vec{r}^{\prime },t\right) V\left( \vec{r}^{\prime }-\vec{r}\right) \hat{\Psi%
}\left( \vec{r}^{\prime },t\right) \right] \hat{\Psi}\left(
\vec{r},t\right).  \label{gp}
\end{align}

By replacing $\hat{\Psi}\left( \vec{r},t\right) $ with the
condensate wave function $\psi $ gives the zeroth-order
approximation to the Heisenberg
equation. In the integral containing the particle-particle interaction $%
V\left( \vec{r}^{\prime }-\vec{r}\right) $ this replacement is in
general a poor approximation for short distances. However, in a
dilute and cold gas, only binary collisions at low energy are
relevant and these collisions are characterized by a single
parameter, the $s$-wave scattering length, independently of the
details of the two-body potential. Therefore, one can replace
$V\left( \vec{r}^{\prime }-\vec{r}\right) $ with an effective
interaction $V\left( \vec{r}^{\prime }-\vec{r}\right) =\lambda
\delta \left( \vec{r}^{\prime }-\vec{r}\right) $, where the
coupling constant $\lambda $
is related to the scattering length $a$ through $\lambda =4\pi \hbar ^{2}a/m$%
. Hence, we assume that in a medium composed of scalar particles
with non-zero mass, the range of Van der Waals-type scalar
mediated interactions among nucleons becomes infinite, when the
medium makes a transition to a Bose-Einstein condensed phase.

With the use of the effective potential the integral in the bracket of Eq. (%
\ref{gp}) gives $\lambda \left| \psi \left( \vec{r},t\right)
\right| ^{2}$, and the resulting equation is the Schrodinger
equation with a quartic nonlinear term \cite{Da99}.

However, in order to obtain a more general description of the
Bose-Einstein condensate stars, we shall assume an arbitrary
non-linear term $g\left( \left| \psi \left( \vec{r},t\right)
\right| ^{2}\right) $ \cite{Ba01}, where we have denoted
\begin{align}
      \rho=\left|\psi\left(\vec{r},t\right)\right|^{2}.
\end{align}

The probability density $\rho $ is normalized according to $\int d^{3}\vec{r}%
\rho =N$. As was pointed out in \cite{Ko00}, the Gross-Pitaevskii
approximation is a long-wavelength theory widely used to describe
a variety of properties of dilute Bose condensates, but for
short-ranged repulsive interactions this theory fails in low
dimensions, and some essential modifications of the theory are
necessary.

Therefore the generalized Gross-Pitaevskii equation describing a
gravitationally trapped rotating Bose-Einstein condensate is given
by
\begin{align}  \label{gen}
 i \hbar \frac{\partial }{\partial
t}\psi \left( \vec{r},t\right) =\left[ -\frac{\hbar
^{2}}{2m}\nabla ^{2}+V_{rot}\left( \vec{r}\right) +V_{ext}\left(
\vec{r}\right) +g^{\prime}\left( \left| \psi \left(
\vec{r},t\right) \right| ^{2}\right) \right] \psi \left(
\vec{r},t\right),
\end{align}
where we denoted $g^{\prime}=dg/d\rho $. The mass of the condensed
particles is denoted by $m$.

As for $V_{ext}\left( \vec{r}\right) $, we assume that it is the
gravitational potential $V$, $V_{ext}=V$, and it satisfies the
Poisson equation
\begin{align}
      \nabla^{2}V=4\pi G\rho _{m},
      \label{pot}
\end{align}
where $\rho _{m}=m\rho $ is the mass density inside the
Bose-Einstein condensate.

The physical properties of a Bose-Einstein condensate described by
the generalized Gross-Pitaevskii equation given by Eq. (\ref{gen})
can be understood much easily by using the so-called Madelung
representation of the wave function \cite{Ba05}, which consist in
writing $\psi $ in the form
\begin{align}
\psi \left( \vec{r},t\right) =\sqrt{\rho \left( \vec{r},t\right)
}\exp \left[ \frac{i}{\hbar }S\left( \vec{r},t\right) \right] ,
\end{align}
where the function $S\left( \vec{r},t\right) $ has the dimensions
of an action.

By substituting the above expression of the wave function into Eq. (%
\ref{gen}) it decouples into a system of two differential
equations for the real functions $\rho $ and $\vec{v}$, given by
\begin{align}
\frac{\partial \rho _{m}}{\partial t}+\nabla \cdot \left( \rho _{m}\vec{v}%
\right) =0,
\end{align}
\begin{align}\label{euler}
\rho _{m}\left[ \frac{\partial \vec{v}}{\partial t}+\left(
\vec{v}\cdot \nabla \right) \vec{v}\right] =-\nabla P\left(
\frac{\rho _{m}}{m}\right) -\rho _{m}\nabla \left(
\frac{V_{rot}}{m}\right) -\rho _{m}\nabla \left(
\frac{V_{ext}}{m}\right) -\nabla \cdot \sigma ^{Q},
\end{align}
where we have introduced the quantum potential
\begin{align}
V_{Q}=-\frac{\hbar ^{2}}{2m}\frac{\nabla ^{2}\sqrt{\rho
}}{\sqrt{\rho }},
\end{align}
the velocity of the quantum fluid
\begin{align}
\vec{v}=\frac{\nabla S}{m},
\end{align}
respectively, and we denoted
\begin{align}
P\left( \frac{\rho _{m}}{m}\right) =g^{\prime }\left( \frac{\rho _{m}}{m}%
\right) \frac{\rho _{m}}{m}-g\left( \frac{\rho _{m}}{m}\right).
\label{state}
\end{align}

From its definition it follows that the velocity field is
irrotational, satisfying the condition $\nabla \times \vec{v}=0$.

The quantum potential $V_{Q}$ has the property \citep{Ba05}
\begin{align}
\rho \nabla _{i}V_{Q}=\nabla _{j}\left( -\frac{\hbar ^{2}}{4m}\rho
\nabla _{i}\nabla _{j}\ln \rho \right) =\nabla _{j}\sigma
_{ij}^{Q},
\end{align}
where $\sigma _{ij}^{Q}=-$ $\hbar ^{2}\rho \nabla _{i}\nabla
_{j}\ln \rho /4m $ is the quantum stress tensor, which has the
dimensions of a pressure and is an intrinsically anisotropic
quantum contribution to the equations of motion.

Therefore the equations of motion of the gravitational ideal
Bose-Einstein condensate take the form of the equation of
continuity and of the hydrodynamic Euler equations. The
Bose-Einstein gravitational condensate can be described as a
gas whose density and pressure are related by a barotropic equation of state %
\citep{Ba01}. The explicit form of this equation depends on the
form of the non-linearity term $g$.

For a static ideal condensate, $\vec{v}\equiv 0$. In this case
from Eq. (\ref {euler}) we obtain
\begin{align}
V_{Q}+V_{rot}+V_{ext}+g^{\prime }={\rm constant} \label{cons2}
\end{align}

By applying the operator $\nabla ^{2}$ to both sides of Eq.
(\ref{cons2}) gives
\begin{align}
\nabla ^{2}\left( V_{Q}+V_{rot}+g^{\prime }\right) +\nabla
^{2}V_{ext}=0.
\end{align}

In the case of a condensate with a non-linearity of the form
$g\left( \rho _{m}\right) =k\rho _{m}^{2}/2$ and in the presence
of a confining
gravitational field $V=V_{ext}$, it follows that the generalized potential $%
V_{gen}=-V_{Q}-V_{rot}-k\rho _{m}$ satisfies the Poisson equation,
\begin{align}
\nabla ^{2}V_{gen}=4\pi G\rho _{m}.
\end{align}

If the quantum potential can be neglected, the mass density of the
condensate is described by the Helmholtz type equation
\begin{align}
\nabla ^{2}\rho _{m}+\frac{4\pi G}{k}\rho _{m}+\frac{\omega
^{2}}{k}=0.
\end{align}

\section{Static and slowly rotating Newtonian Bose-Einstein condensates }

When the number of particles in the gravitationally bounded
Bose-Einstein condensate becomes large enough, the quantum
pressure term makes a significant contribution only near the
boundary of the condensate. Hence it is much smaller than the
non-linear interaction term. Thus the quantum stress term in the
equation of motion of the condensate can be neglected. This is the
Thomas-Fermi approximation, which has been extensively used for
the study of the Bose-Einstein condensates \cite{Da99}. As the
number of particles in the condensate becomes infinite, the
Thomas-Fermi approximation becomes exact \cite{Wa01}. This
approximation also corresponds to the classical limit of the
theory (it corresponds to neglecting all terms with powers of
$\hbar$ or as the regime of strong repulsive interactions among
particles. From a mathematical point of view the Thomas-Fermi
approximation corresponds to neglecting in the equation of motion
all terms containing $\nabla{\rho}$ and $\nabla{S}$.

In the case of a static Bose-Einstein condensate, all physical
quantities are independent of time. Moreover, in the first
approximation we also neglect the rotation of the condensate,
taking $V_{rot}=0$. Therefore the equations describing the static
Bose-Einstein condensate in a gravitational field with potential
$V$ take the form
\begin{align}
\nabla P\left( \frac{\rho _{m}}{m}\right) =-\rho _{m}\nabla
\left(\frac{V}{m}\right) ,
\end{align}
\begin{align}
\nabla ^{2}V=4\pi G\rho _{m}.
\end{align}

These equations must be integrated together with the equation of state $%
P=P\left( \rho _{m}\right) $, which follows from Eq.
(\ref{state}), and some appropriately chosen boundary conditions.
By assuming that the non-linearity in the Gross-Pitaevskii
equation is of the form
\begin{align}
g\left( \rho \right) =\alpha \rho ^{\Gamma },
\end{align}
where $\alpha $ and $\Gamma $ are positive constants, $\alpha =\mathrm{%
constant}>0$, $\Gamma =\mathrm{constant}>0$, it follows that the
equation of state of the gravitational Bose-Einstein condensate is
the polytropic equation of state,
\begin{align}
P\left( \rho _{m}\right) =\alpha \left( \Gamma -1\right) \rho
_{m}^{\Gamma }=K\rho _{m}^{\Gamma },
\end{align}
where we denoted $K=\alpha \left( \Gamma -1\right) $.

By representing $\Gamma $ in the form $\Gamma =1+1/n$, where $n$
is the polytropic index, it follows that the structure of the
static gravitationally bounded Bose-Einstein condensate is
described by the Lane-Emden equation,
\begin{align}
      \frac{1}{\xi^{2}}\frac{d}{d\xi}
      \left(\xi^{2}\frac{d\theta}{d\xi}\right) + \theta^{n} = 0,
      \label{laneemden}
\end{align}
where $\theta $ is a dimensionless variable defined via $\rho
_{m}=\rho _{cm}\theta ^{n}$ and $\xi $ is a dimensionless
coordinate introduced via the transformation $r=\left[ (n+1)K\rho
_{cm}^{1/n-1}/4\pi G\right]
^{1/2}\xi $. $\rho _{cm}$ is the central density of the condensate %
\cite{Ch57}.

Hence the mass and the radius of the condensate are given by
\begin{align}
R=\left[ \frac{\left( n+1\right) \alpha }{4\pi Gn}\right]
^{1/2}\rho _{cm}^{(1-n)/2}\xi _{1},
\end{align}
and
\begin{align}
M=4\pi \left[ \frac{\left( n+1\right) \alpha }{4\pi Gn}\right]
^{3/2}\rho _{cm}^{(3-n)/2n}\xi _{1}^{2}\left| \theta ^{\prime
}\left( \xi _{1}\right) \right| ,
\end{align}
respectively, where $\xi _{1}$ defines the zero- pressure and
zero-density surface of the condensate, $\theta \left( \xi
_{1}\right) =0$ \cite{Ch57}.

In the standard approach to the Bose-Einstein condensates, the
non-linearity term $g$ is given by
\begin{align}
g\left( \rho \right) =\frac{u_{0}}{2}\left| \psi \right| ^{4}=\frac{u_{0}}{2}%
\rho ^{2},
\end{align}
where $u_{0}=4\pi \hbar ^{2}a/m$ \cite{Da99}. The corresponding
equation of state of the condensate is
\begin{align}
P\left( \rho _{m}\right) =U_{0}\rho _{m}^{2},
\end{align}
with
\begin{align}
      U_{0} = \frac{2\pi \hbar^{2} a}{m^{3}}.
\end{align}

Therefore the equation of state of the Bose-Einstein condensate
with quartic non-linearity is a polytrope with index $n=1$. In
this case the solution of the Lane-Emden equation can be obtained
in an analytical form, and the solution satisfying the boundary
condition $\theta \left( 0\right) =1$ is \cite{Ch57}
\begin{align}
      \label{theta}
      \theta(\xi) = \frac{\sin\xi}{\xi}.
\end{align}

The radius of the gravitationally bounded system is defined by the
condition $\theta \left( \xi _{1}\right) =0$, giving $\xi _1=\pi
$. Therefore the radius $R$ of the Bose-Einstein condensate is
given by
\begin{align}\label{rad}
R=\pi \sqrt{\frac{\hbar ^{2}a}{Gm^{3}}}.
\end{align}

The radius of the gravitationally bounded Bose-Einstein condensate
with quartic non-linearity is independent of the central density
and the mass of the system, and depends only on the physical
characteristics of the condensate.

The total mass of the condensate is obtained as
\begin{align}
M=4\pi ^{3}\left( \frac{\hbar ^{2}a}{Gm^{3}}\right) ^{3/2}\rho
_{cm}\left| \theta ^{\prime }\left( \xi _{1}\right) \right| =4\pi
^{2}\left( \frac{\hbar ^{2}a}{Gm^{3}}\right) ^{3/2}\rho _{cm},
\end{align}
where we have used $\left| \theta ^{\prime }\left( \xi _{1}\right)
\right| =1/\pi $.

The mass-radius relation for the static condensate is given by
\begin{align}\label{mass0}
M=\frac{4}{\pi }R^{3}\rho _{cm},
\end{align}
which shows that the mean density of the condensate $\left\langle
\rho _{m}\right\rangle =3M/4\pi R^{3}$ can be obtained from the
central density of the condensate by the relation
\begin{align}\label{mean}
\left\langle \rho _{m}\right\rangle =3\frac{\rho _{cm}}{\pi ^{2}}.
\end{align}

The case of the slowly rotating Bose-Einstein condensates can also
be straightforwardly analyzed, by taking into account the fact
that the condensate obeys a polytropic equation of state. The
study of the slowly rotating polytropes was performed in detail by
Chandrasekhar  \cite{Ch33}.

The Lane-Emden equation for a rotating Bose-Einstein condensate is
\begin{align}
\frac{1}{\xi ^{2}}\frac{\partial }{\partial \xi }\left( \xi ^{2}\frac{%
\partial \theta }{\partial \xi }\right) +\frac{1}{\xi ^{2}}\frac{\partial }{%
\partial \mu }\left[ \left( 1-\mu ^{2}\right) \frac{\partial \theta }{%
\partial \mu }\right] =-\theta ^{n}+\Omega ,
\end{align}
where $\mu =\cos \theta $ and $\Omega =\omega ^{2}/2\pi G\rho
_{cm}$. The radius $R_{\omega }$ and the mass $M_{\omega }$ of the
condensate in slow rotation are given in the first order in
$\Omega $ by
\begin{align}
R_{\omega }=\left[ \frac{\left( n+1\right) \alpha }{4\pi
Gn}\right] ^{1/2}\rho _{cm}^{(1-n)/2}\xi _{1}\left[ 1+\frac{3\psi
_{0}\left( \xi _{1}\right) }{\xi _{1}\left| \theta ^{\prime
}\left( \xi _{1}\right) \right| }\Omega \right] ^{1/3},
\end{align}
and
\begin{align}
M_{\omega }=4\pi \left[ \frac{\left( n+1\right) \alpha }{4\pi
Gn}\right] ^{3/2}\rho _{cm}^{(3-n)/2n}\xi _{1}^{2}\left| \theta
^{\prime }\left( \xi _{1}\right) \right| \left[ 1+\frac{\xi
_{1}/3-\psi _{0}^{^{\prime }}\left(
\xi _{1}\right) }{\left| \theta ^{\prime }\left( \xi _{1}\right) \right| }%
\Omega \right] ,
\end{align}
respectively. The values of the function $\psi _{0}$ are tabulated
in \cite {Ch33}.

In the case of the Bose-Einstein condensate with quartic
non-linearity, with polytropic index $n=1$, the Lane-Emden
equation can be integrated exactly, giving for the radius and mass
of the rotating condensate the following simple relations
\begin{align}
R_{\omega }=R\left( 1+3\Omega \right) ^{1/3},
\end{align}
\begin{align}
M_{\omega }=M\left[ 1+\left( \frac{\pi ^{2}}{3}-1\right) \Omega
\right].
\end{align}

With respect to a scaling of the parameters $m$, $a$ and $\rho
_{cm}$ of the form $m\rightarrow \alpha _{1}m$, $a\rightarrow
\alpha _{2}a$, $\rho _{cm}\rightarrow \alpha _{3}\rho _{cm}$, the
radius and the mass of the condensate have the following scaling
properties:
\begin{align}\label{scal}
R\rightarrow \alpha _{1}^{-3/2}\alpha _{2}^{1/2}R,M\rightarrow
\alpha _{1}^{-9/2}\alpha _{2}^{3/2}\alpha _{3}M.
\end{align}

\section{Dark matter as a Bose-Einstein condensate}

In the present Section we analyze, by using the results obtained
in the previous Sections, the possibility that dark matter is in
the form of a Bose-Einstein condensate. For simplicity we shall
restrict our study to the case of a Bose-Einstein condensate with
a quartic non-linearity. As a first result one has to point out
that, if dark matter is a Bose-Einstein condensate, it cannot be
pressure-less, as usually considered in most of the investigations
of the galactic dark matter problem, but must obey a polytropic
equation of state with index $n=1$.

The possibility that dark matter has a substantial amounts of
pressure, comparable in magnitude to the energy density, has been
discussed in \cite{BhKa03}. Galaxy halo models, consistent with
observations of flat rotation curves, are possible for a variety
of equations of state with anisotropic pressures. However, in the
case of dark matter in the form of the Bose-Einstein condensate
the pressure distribution is isotropic.

The density distribution of the dark matter Bose-Einstein
condensate follows from Eq. (\ref{theta}) and is given by
\begin{align}\label{dens}
\rho _{DM}\left( r\right) =\rho _{DM}^{(c)}\frac{\sin kr}{kr},
\end{align}
where $k=\sqrt{Gm^{3}/\hbar ^{2}a}$ and $\rho _{DM}^{(c)}$ is the
central density of the condensate, $\rho _{DM}^{(c)}=\rho
_{DM}\left( 0\right) $. The mass profile of the dark condensate
galactic halo $m_{DM}\left( r\right) =4\pi \int_{0}^{r}\rho
_{DM}\left( r\right) r^{2}dr$ is
\begin{align}
      m_{DM}\left( r\right) =\frac{4\pi \rho _{DM}^{(c)}}{k^{2}}r\left(
      \frac{\sin kr}{kr}-\cos kr\right).
      \label{mass}
\end{align}

Eq. (\ref{mass}) allows to represent the tangential velocity $%
v_{tg}^{2}\left( r\right) =Gm_{DM}(r)/r$ of a test particle moving
in the dark halo as
\begin{align}
      v_{tg}^{2}\left( r\right) =\frac{4\pi G\rho _{DM}^{(c)}}{k^{2}}
      \left( \frac{\sin kr}{kr}-\cos kr\right).
      \label{vel}
\end{align}
For $r\rightarrow 0$ we have $v_{tg}^{2}\left( r\right)
\rightarrow 0$.

At the boundary of the halo, which is defined by the radius
$R_{DM}$ of the halo, the density of the Bose-Einstein condensate
is negligible, $\rho _{DM}(R_{DM})=0$ and $kr\rightarrow \pi $.
These conditions give the radius of the dark matter condensate as
\begin{align}
R_{DM}=\pi \sqrt{\frac{\hbar ^{2}a}{Gm^{3}}}.
\end{align}

The total mass of the condensate is given by
\begin{align}
M_{DM}=m_{DM}\left( R_{DM}\right) =\frac{4}{\pi }R_{DM}^{3}\rho
_{DM}^{(c)}.
\end{align}

Near the vacuum boundary of the condensate the maximum tangential
velocity of a test particle tends to a constant value which can be
expressed as
\begin{align}\label{vel1}
 v_{tg}^{2}\left( R\right) =
\frac{4\pi G\rho _{DM}^{(c)}}{k^{2}}=\frac{4G\rho
_{DM}^{(c)}R_{DM}^{2}}{\pi }.
\end{align}

Eqs. (\ref{dens})-(\ref{vel1}) give a complete description of the
physical properties of the galactic dark matter condensates in
terms of three quantities, the mass $m$ of the condensed particle,
the inter-particle scattering length $a$, and the central density
of the condensate. Together with Eqs. (\ref{rad}) and
(\ref{mass0}) they can be used to constrain the physical
properties of the condensate.

As a toy-model we consider the example of a galactic dark matter
halo extending to up to $R_{DM}=10$ kpc $\approx 3.08\times
10^{22}$ cm, with a mass of the order of $M_{DM}=3\times
10^{11}M_{\odot}$. The mean density of the condensate dark matter
is $\left\langle \rho _{DM}\right\rangle =3M_{DM}/4\pi
R_{DM}^{3}\approx 5.30\times 10^{-24}$ g/cm$^{3}$. Then Eq.
(\ref{mean}) gives the central density of the condensate as $\rho
_{DM}^{(c)}=\pi ^{2}\left\langle \rho _{DM}\right\rangle /3\approx
1.74\times 10^{-23}$ g/cm$^{3}$.

An important parameter in the description of the Bose-Einstein
condensate is the scattering length $a$. In terrestrial
experiments performed with $^{87}$Rb $a$ has a value of the order
of $a\left(^{87}{\rm Rb}\right)\approx 5.77\times 10^{-7}\;{\rm
cm}\approx 5.77\times 10^6\;{\rm fm}$ \cite{exp}. However, for
very light particles $a$ may be much larger.

The mass of the particle in the condensate can be obtained from
the radius of the dark matter halo as
\begin{align}
      m = \left(\frac{\pi^{2}\hbar^{2}a}{GR^{2}}\right)^{1/3}\approx
      2. 58\times 10^{-30}\left[ a\left( {\rm cm}\right) \right]
      ^{1/3}\left[R\;{\rm (kpc)}\right]^{-2/3}{\rm g}
      \approx 6.73 \times 10^{-2}\left[ a\left( {\rm fm}\right) \right]^{1/3}
      \left[R\;{\rm (kpc)}\right]^{-2/3} {\rm eV}.
\end{align}

From this equation it follows that the mass of the particle
forming the condensate dark matter halos is of the order of eV.
For $a\approx 1$ fm and $R\approx 10$ kpc, the mass may be of the
order of $m\approx 14$ meV. For values of $a$ of the order of
$a\approx 10^6$ fm, corresponding to the values of $a$ observed in
terrestrial laboratory experiments, $m\approx 1.44$ eV. Of course
much larger values of $a$ may bring the mass of the condensate
particle in the range of a few tenth of eV's.

With the use of Eq. (\ref{vel1}) one can estimate the tangential
velocity of a test particle in the dark matter halo. By using the
numerical values from the previous estimations we obtain
$v_{tg}(R_{DM})\approx 365$ km/s, a value which is consistent with
the observations of the galactic rotation curves. From the
definition of the mean density it follows that the total mass of
the condensate can be obtained from the tangential velocity as
$M_{DM}\approx v_{tg}^{2}(R_{DM})R_{DM}/G$.

\begin{figure}[!ht]
\centering
\includegraphics[width=0.48\linewidth]{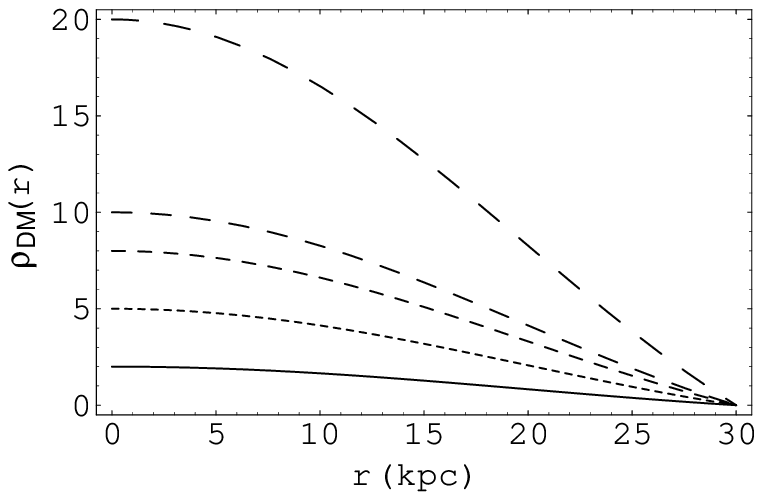}
\includegraphics[width=0.48\linewidth]{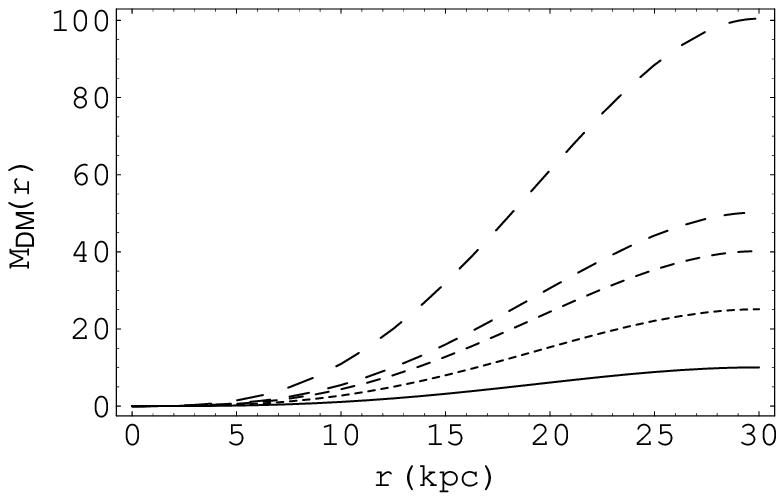}\\[1ex]
\caption{Density profile (in units of $10^{-25}$ g/cm$^{3}$) (left
figure) and mass distribution (in units of $10^{10}M_{\odot }$)
(right figure) of the Newtonian Bose-Einstein condensate dark
matter for different values of the central density: $\rho
_{DM}^{(c)}=2\times 10^{-25}$ g/cm$^3$ (solid curve), $\rho
_{DM}^{(c)}=5\times 10^{-25}$ g/cm$^3$ (dotted curve), $\rho
_{DM}^{(c)}=8\times 10^{-25}$ g/cm$^3$ (short dashed curve), $\rho
_{DM}^{(c)}=\times 10^{-24}$ g/cm$^3$ (long dashed curve) and
$\rho _{DM}^{(c)}=2\times 10^{-24}$ g/cm$^3$ (ultra-long dashed
curve).} \label{FIG1}
\end{figure}

The density profile of the Newtonian dark matter condensate and its
mass profile are represented in Fig.~\ref{FIG1}.

\begin{figure}[!ht]
\includegraphics[width=0.48\linewidth]{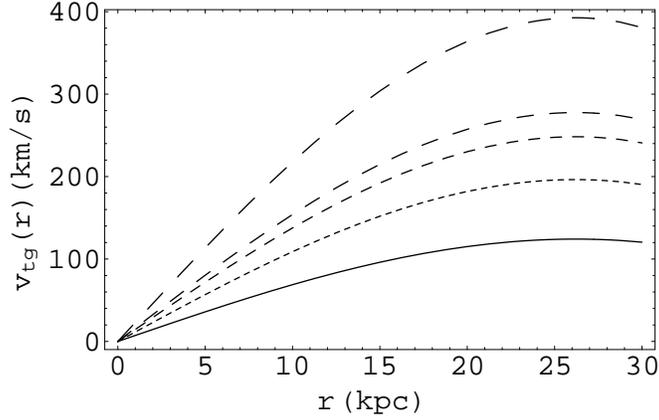}
\caption{Tangential velocity of a test particle in a galactic
Newtonian Bose-Einstein condensate for different values of the
central density: $\rho _{DM}^{(c)}=2\times 10^{-25}$ g/cm$^3$
(solid curve), $\rho _{DM}^{(c)}=5\times 10^{-25}$ g/cm$^3$
(dotted curve), $\rho _{DM}^{(c)}=8\times 10^{-25}$ g/cm$^3$
(short dashed curve), $\rho _{DM}^{(c)}=\times 10^{-24}$ g/cm$^3$
(long dashed curve) and $\rho _{DM}^{(c)}=2\times 10^{-24}$
g/cm$^3$ (ultra-long dashed curve).} \label{FIG3}
\end{figure}

The tangential velocity of a test particle in the gravitationally
bounded galactic condensate is shown, for different values of the
central density, in Fig.~\ref{FIG3}.

\section{The general relativistic Bose-Einstein condensate galactic halos}

In the previous Sections we have considered the gravitationally
bounded Bose-Einstein condensate dark matter in the framework of
Newtonian gravity. General relativistic effects may change the
physical properties of the dark matter condensate halos in both a
qualitative and quantitative way. These effects may become
particularly important since as a condensate dark matter cannot be
described as a pressure-less fluid. Therefore the study of the
general relativistic Bose-Einstein condensates offers a better
understanding of their physical properties, and opens the
possibility of direct observational testing of their properties.

For a static, spherically symmetric distribution of matter in the
galactic halo the line element is
\begin{align}
      \label{line}
      ds^{2} = -e^{\nu(r)} c^2 dt^{2} + e^{\lambda(r)} dr^{2} + r^{2}
      \left(d\theta^{2} + \sin^{2}\theta d\phi^{2}\right).
\end{align}

The galactic rotation curves provide the most direct method of
analyzing the gravitational field inside a spiral galaxy. The
rotation curves have been determined for a great number of spiral
galaxies. They are obtained by measuring the frequency shifts $z$
of the light emitted from stars and from the 21-cm radiation
emission from the neutral gas clouds \cite{Bi87}. As shown in
Appendix~\ref{appI}, the tangential velocity of test particles in stable
circular orbits is given by \cite{La03}
\begin{align}
\frac{v_{tg}^{2}}{c^2}=\frac{r\nu ^{\prime }}{2}.
\end{align}
Thus, the rotational velocity of the test body is determined by
the metric coefficient $\exp \left( \nu \right) $ only.

A general relativistic static dark matter distribution with energy density
$\rho _{DM}(r)$ and pressure $P_{DM}(r)$ can be described by the mass continuity
equation and by the Tolman-Oppenheimer-Volkoff (TOV) equation, which are given
by \citep{Gl00}
\begin{align}
      \frac{dm_{DM}}{dr}=4\pi \rho _{DM}r^2 ,
      \label{s1}
\end{align}
\begin{align}
      \frac{dP_{DM}(r)}{dr}=-\frac{\left( G/c^{2}\right)
      \left(\rho_{DM}c^{2}+P_{DM}\right)
      \left(4\pi P_{DM}r^{3}/c^{2}+m_{DM}\right)}
      {r^{2}\left[ 1-\frac{2Gm_{DM}(r)}{c^{2}r}\right]},
      \label{s2}
\end{align}
\begin{align}
      \label{s3}
      \frac{d\nu }{dr}=-\frac{2P_{DM}^{\prime }(r)}{\rho_{DM}c^{2}+P_{DM}} =
      \frac{2\left(G/c^{2}\right) \left( 4\pi P_{DM}r^{3}/c^{2}+m_{DM}\right)}
      {r^{2}\left[ 1-\frac{2Gm_{DM}(r)}{c^{2}r}\right]},
\end{align}
where $m_{DM}(r)$ is the mass inside radius $r$. The system of
equations (\ref{s1})-(\ref{s2}) must be closed by choosing the
equation of state for the thermodynamic pressure of the dark
matter,
\begin{align}
      P_{DM}=P_{DM}\left(\rho_{DM}\right).
\end{align}

At the center of the dark matter distribution the mass must
satisfy the boundary condition
\begin{align}
      m_{DM}(0)=0.
\end{align}
For the thermodynamic pressure of the dark matter $P_{DM}$ we
assume that it vanishes on the surface of the dark halo, $P_{DM}(R)=0$.
The exterior of the Bose-Einstein condensate halo is characterized
by the Schwarzschild metric, describing the vacuum outside the
galaxy, and given by \citep{Gl00}
\begin{align}
      \left( e^{\nu }\right)^{ext} = \left( e^{-\lambda }\right)^{ext} =
      1-\frac{2GM_{DM}}{r},\qquad r\geq R,
\end{align}
where $M_{DM}=m_{DM}(R)$ is the total mass of the dark halo. The
interior solution must match with the exterior solution on the
vacuum boundary of the star.

As for the equation of state of the dark matter we adopt the
equation of state corresponding to the Bose-Einstein condensate
with quartic nonlinearity, given by
\begin{align}
      P_{DM} \left(\rho_{DM}\right) = U_{0}\rho_{DM}^{2},
\end{align}
with $U_{0}=$ $2\pi \hbar ^{2}a/m^{3}$.

The structure equations for the Bose-Einstein condensate stars can
be written in a dimensionless form, by introducing a set of dimensionless variables $%
\eta $, $M_{0}$, $\theta $ and $\Sigma $, respectively, and
defined as
\begin{align}
      \label{var}
      r=r^{\ast}\eta,\quad m_{DM}=m^{\ast }M_{0},\quad
      \rho_{DM}=\rho_{DM}^{(c)}\theta,\quad P_{DM}=\rho_{DM}^{(c)}c^{2}\lambda \theta^2,
\end{align}
where $\rho _{DM}^c$ is the central ($r=0$) value of the energy
density of the Bose-Einstein condensate. By taking
\begin{align}
      r^{\ast}=\frac{c}{\sqrt{4\pi G\rho _{DM}^{(c)}}},
\end{align}
and
\begin{align}
m^{\ast }=4\pi \left(r^{\ast}\right)^{3}\rho
_{DM}^{(c)}=\frac{c^3}{\sqrt{4\pi G^{3}\rho _{DM}^{(c)}}},
\end{align}
respectively, and by denoting
\begin{align}
\lambda _0=\frac{U_{0}\rho _{DM}^{(c)}}{c^{2}},
\end{align}
the structure equations of the Bose-Einstein condensate galactic
dark matter halos with quartic non-linearity can be written as
\begin{align}\label{s4}
\frac{dM_{0}}{d\eta }=\theta \eta ^{2},
\end{align}
\begin{align}\label{s5}
\frac{d\theta }{d\eta }=-\frac{\left( 1+\lambda _0\theta \right)
\left( \lambda _0\theta ^{2}\eta ^{3}+M_{0}\right) }{2\lambda
_0\eta ^{2}\left( 1-\frac{2M_0}{\eta }\right) },
\end{align}
\begin{align}\label{s6}
\frac{d\nu }{d\eta }=-\frac{4\lambda _0}{1+\lambda _0\theta }\frac{d\theta }{%
d\eta }=\frac{2\left( \lambda _0\theta ^{2}\eta ^{3}+M_{0}\right)
}{\eta ^{2}\left( 1-\frac{2M_0}{\eta }\right) }.
\end{align}

In the new dimensionless variable the boundary conditions at the
center and surface of the condensate halos are given by
\begin{align}
      M_{0}(0)=0,\quad \theta(0)=1,\quad \theta(\eta_{S})=0,
\end{align}
where $\eta_{S}$ is the value of the dimensionless radial
coordinate at the vacuum boundary of the galaxy.

\begin{figure}[!ht]
\includegraphics[width=0.48\linewidth]{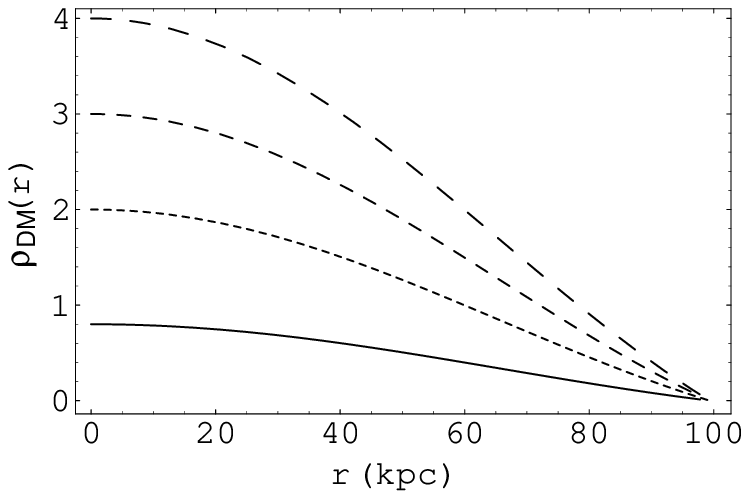}
\includegraphics[width=0.48\linewidth]{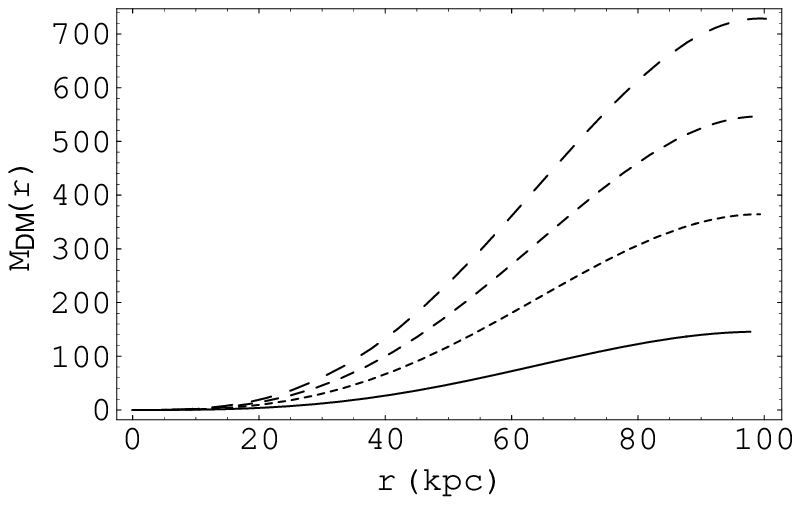}\\[1ex]
\caption{General relativistic density profiles (in units of
$10^{-25}$ g/cm$^3$) (left figure) and mass profiles (in units of
$10^{10}M_{\odot }$) (right figure) for a Bose-Einstein condensate
galactic halo with $m=5.6\times 10^{-34}$ g, $a=10^6$ fm and
different values of the central density:
$\rho_{DM}^{(c)}=10^{-25}$ g/cm$^3$ (solid curve),
$\rho_{DM}^{(c)}=2\times 10^{-25}$ g/cm$^3$ (dotted curve),
$\rho_{DM}^{(c)}=3\times 10^{-25}$ g/cm$^3$ (short dashed curve)
and $\rho _{DM}^{(c)}=4\times 10^{-25}$ g/cm$^3$ (long dashed
curve).} \label{FIG4}
\end{figure}

The general relativistic density and mass profiles of the
Bose-Einstein condensate dark halos are represented, for a fixed
value of $m$ and $a$ and for different values of the central
density, in Fig.~\ref{FIG4}, respectively.

\begin{figure}[!ht]
\includegraphics[width=0.48\linewidth]{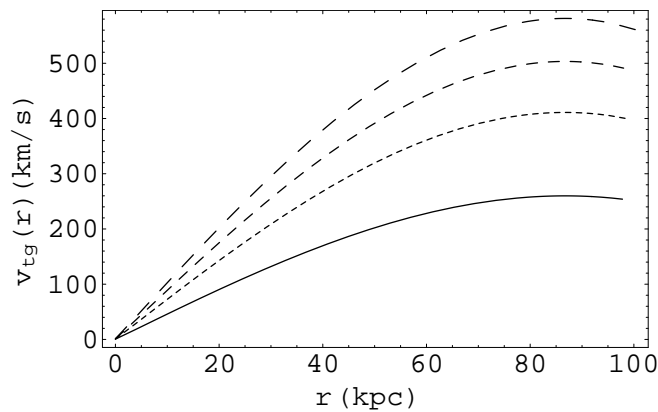}
\caption{Tangential velocity of a test particle in a general
relativistic Bose-Einstein condensate galactic halo with
$m=5.6\times 10^{-34}$ g, $a=10^6$ fm for different values of the
central density: $\rho_{DM}^{(c)}=10^{-25}$ g/cm$^3$ (solid
curve), $\rho_{DM}^{(c)}=2\times 10^{-25}$ g/cm$^3$ (dotted
curve), $\rho_{DM}^{(c)}=4\times 10^{-25}$ g/cm$^3$ (short dashed
curve) and $\rho _{DM}^{(c)}=6\times 10^{-25}$ g/cm$^3$ (long
dashed curve).} \label{FIG6}
\end{figure}

The tangential velocity of a test particle moving in the dark halo
is given by
\begin{align}
v_{tg}=c\sqrt{\frac{\lambda _0\theta ^{2}\eta ^{3}+M_{0}}{\eta \left( 1-\frac{%
2M_{0}}{\eta }\right) }},
\end{align}
and is represented as a function of the distance to the galactic center $r$ in
Fig.~\ref{FIG6}.

For a galactic dark matter halo with mass of the order of
$M_{DM}=10^{12}M_{\odot }$ and radius $R_{DM}=100$ kpc, the quantity $2GM_{DM}/c^{2}R_{DM}$ is of the order of $%
9.6\times 10^{-7}$, which is much smaller than
one. Therefore in the nominator of Eq. (\ref{s2}) one can neglect the term $%
2Gm_{DM}(r)/c^{2}r$. For central densities of the galactic halos of the order of $%
10^{-26}$ g/cm$^{3}$ and for $a=10^{6}$ fm, the quantity $\lambda
_0 =U_{0}\rho _{DM}^{(c)}/c^{2}\approx 10^{-7}$, which implies
$\lambda _0 \theta \ll 1$.

Hence for galactic halos consisting of Bose-Einstein condensates,
we can obtain a generalized Lane-Emden equation that contains the
$1/c^2$ corrections
\begin{align}
      \frac{1}{\eta ^2}\frac{d}{d\eta }\left(\eta ^2\frac{d\theta}{d\eta }\right)
      +\theta = \lambda_0 \theta \left[-4\theta + 6\theta\frac{d\theta}{d\eta}
      + 5\frac{1}{\theta }\left(\frac{d\theta}{d\eta }\right)^2\right].
\label{GLE}
\end{align}

In the zeroth order of approximation ($\lambda_0 \rightarrow 0$
which is equivalent to $c^2 \rightarrow \infty$) the resulting
Lane-Emden equation reduces to the Newtonian limit
Eq.~(\ref{laneemden}) with $n=1$, whose solution is given by
Eq.~(\ref{theta}). Like in the Newtonian case one can obtain a
series expansion of the generalized Lane-Emden equation.

\section{Comparing the model with the observational data}

In order to test our results we compare the predictions of our
model with the observational data on the galactic rotation curves,
obtained for a sample of low surface luminosity and dwarf galaxies
in \cite{deBl02} and \cite{Spek05}, respectively. Generally, in a
realistic situation, a galaxy consists of a distribution baryonic
(normal) matter, consisting of stars of
mass $M_{star}$, ionized gas of mass $M_{gas}$, neutral hydrogen of mass $%
M_{HI}$ etc., and the dark matter of mass $M_{DM}$, which we
assume to be in the form of a Bose-Einstein condensate. The total
mass of the galaxy is therefore
$M_{tot}=M_{star}+M_{gas}+M_{HI}+M_{DM}+\ldots =M_{B}^{tot}+M_{DM}$,
where $M_{B}^{tot}=M_{star}+M_{gas}+M_{HI}+\ldots$ is the total
baryonic mass in the galaxy.

As for the baryonic mass, we assume it is concentrated into an
inner core of radius $r_{c}$, and that its mass profile $M_{B}(r)$
can be described by the simple relation
\begin{align}
M_{B}\left( r\right) =M_{B}^{tot}\left( \frac{r}{r+r_{c}}\right)
^{3\beta },
\end{align}
where $\beta =1$ for high surface brightness galaxies (HSB) and
$\beta =2$ for low surface brightness (LSB) and dwarf galaxies,
respectively \cite{Mo96}. Therefore the tangential velocity of a
test particle in our model, in which dark matter is in a form of a
Bose-Einstein condensate, is given by
\begin{multline}\label{eqexp}
      v_{tg}^{2}\left( {\rm km/s}\right) = 4.45\times 10^{4}
      \frac{M_{B}^{tot}\left( 10^{10}M_{\odot }\right) }{r({\rm kpc})}
      \left( \frac{r}{r+r_{c}}\right) ^{3\beta }\\ +
      8.06\times \rho _{DM}^{(c)}\left( 10^{-25}{\rm g/cm}^{3}\right)
      \times \left( R\;{\rm kpc}\right)^{2}\times
      \left[ \frac{\sin \left( \pi r/R\right) }{\pi r/R}-\cos
      \left( \frac{\pi r}{R}\right) \right] ,
\end{multline}

In order to test the validity of Eq. (\ref{eqexp}), which
represents the basic prediction of our model, we have performed
$12$ galaxy rotation curves fits to a sample of low surface
luminosity (LSB) galaxies and dwarf galaxies, respectively. The
observational data have been taken from \cite{deBl02} for the LSB
galaxies and from \cite{Spek05} for the dwarf galaxies. The fitted
rotation curves are represented by a solid curve in
Fig.~\ref{ddo189-ngc4395}, respectively, where the points
represent the observational data. In all the considered cases we
have adopted for $\beta $ the value $\beta =2$. The numerical
results and the obtained values of the fitting parameters are
summarized in Table~\ref{numtable}.

\begin{figure*}[!h]
\centering\mbox{}\\[-1cm]
\includegraphics[totalheight=0.35\textheight,width=.48\textwidth]{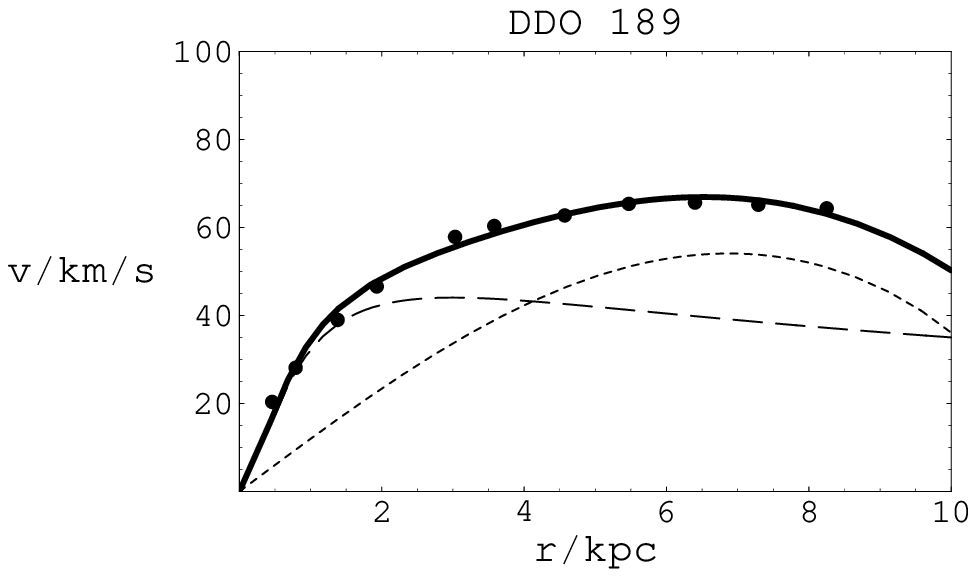}
\includegraphics[totalheight=0.35\textheight,width=.48\textwidth]{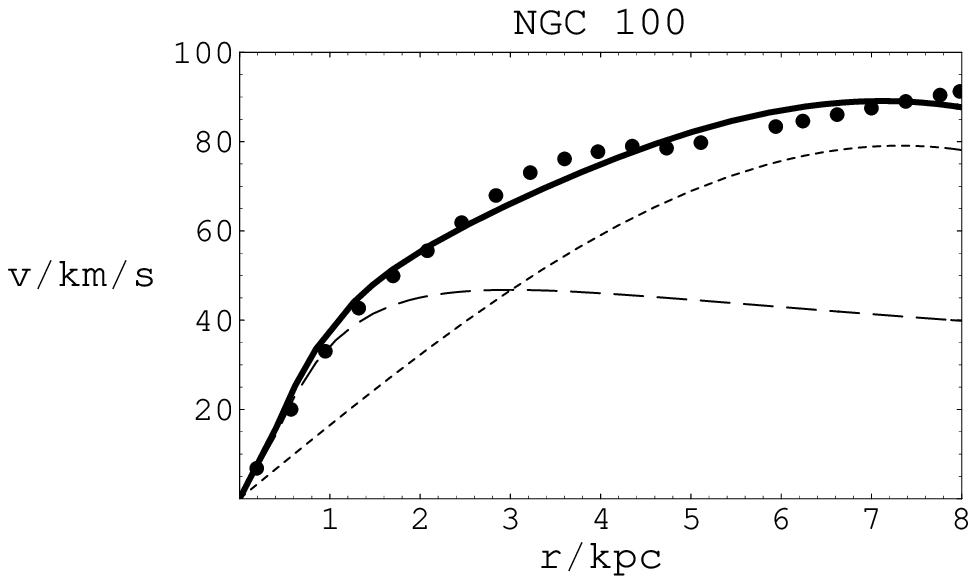}\\[-3cm]
\includegraphics[totalheight=0.35\textheight,width=.48\textwidth]{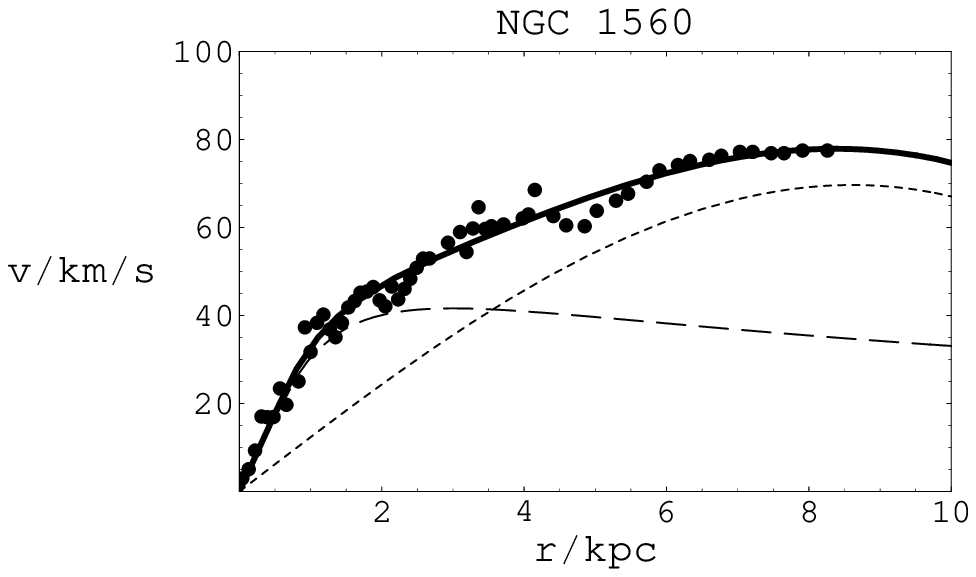}
\includegraphics[totalheight=0.35\textheight,width=.48\textwidth]{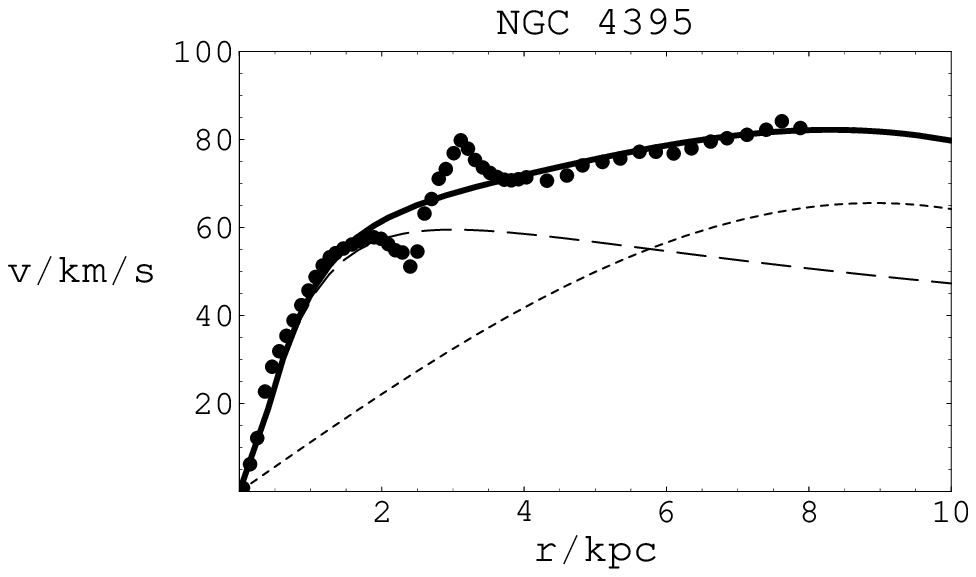}\\[-3cm]
\includegraphics[totalheight=0.35\textheight,width=.48\textwidth]{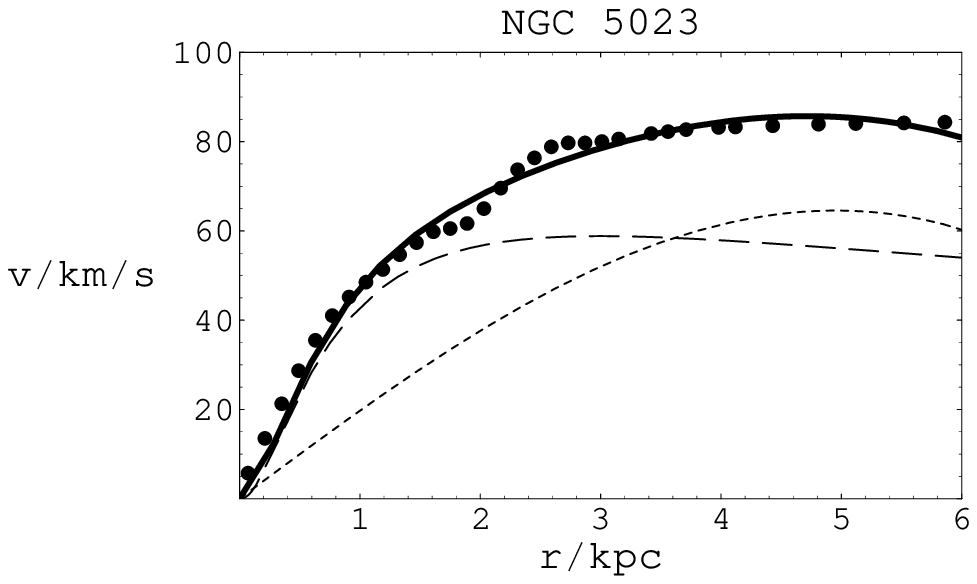}
\includegraphics[totalheight=0.35\textheight,width=.48\textwidth]{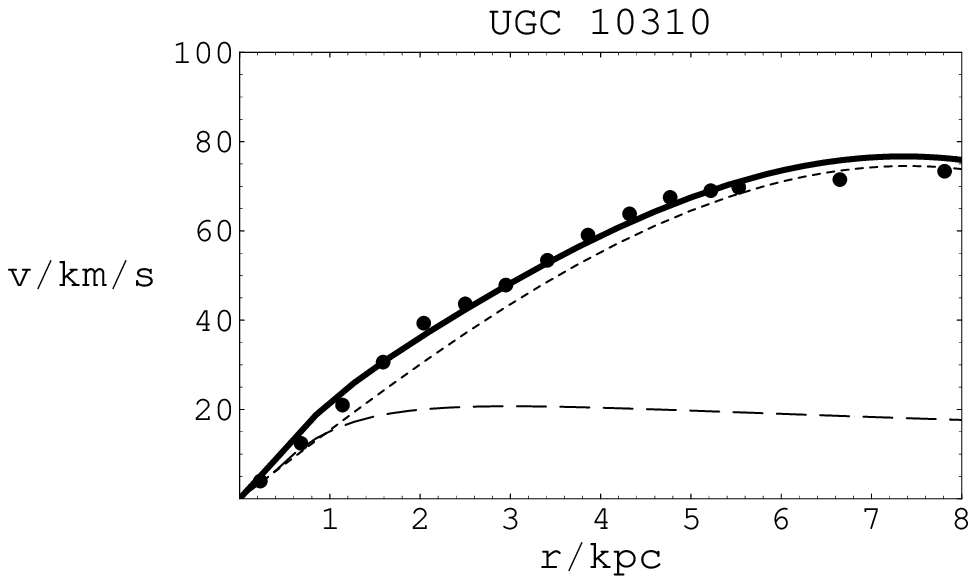}\\[-3cm]
\includegraphics[totalheight=0.35\textheight,width=.48\textwidth]{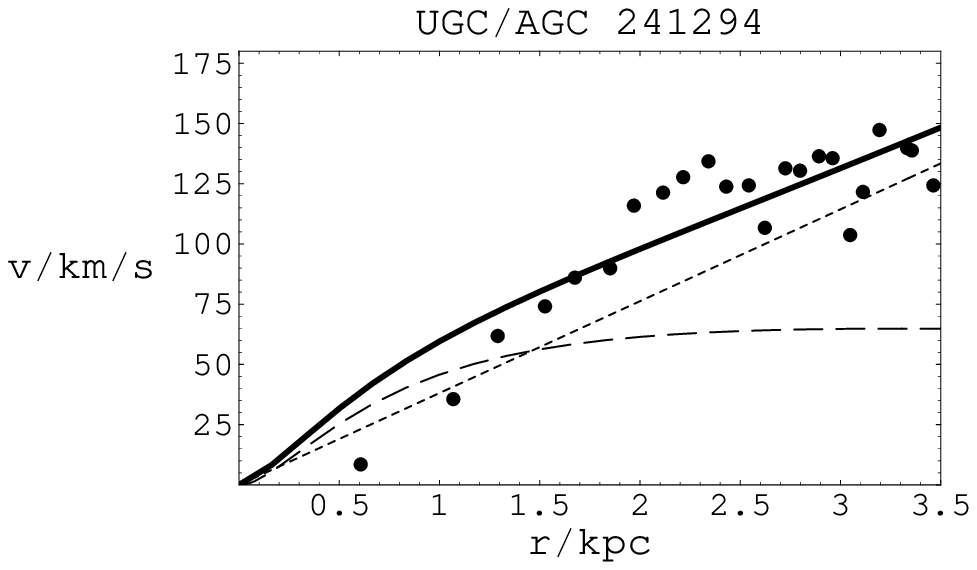}
\includegraphics[totalheight=0.35\textheight,width=.48\textwidth]{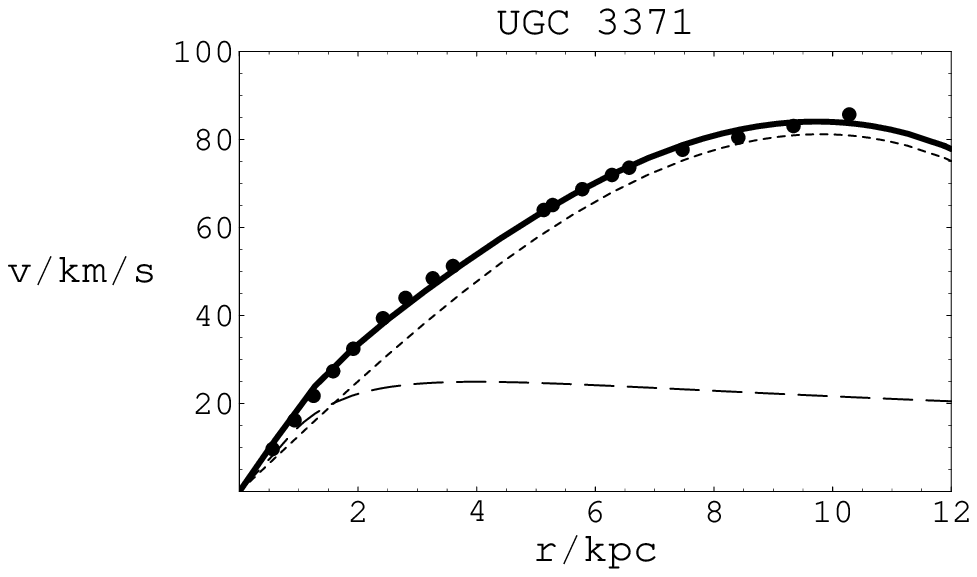}\\[-3cm]
\includegraphics[totalheight=0.35\textheight,width=.48\textwidth]{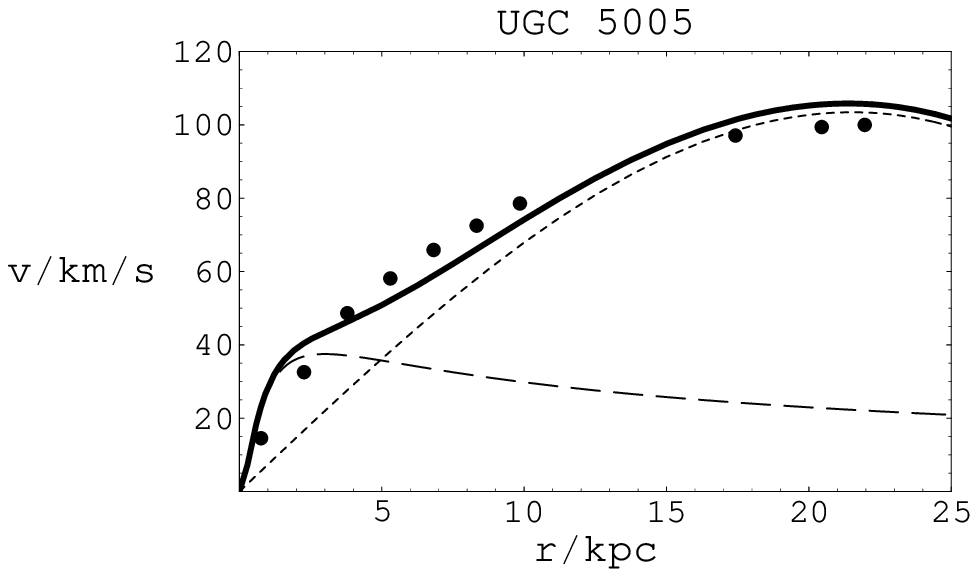}
\includegraphics[totalheight=0.35\textheight,width=.48\textwidth]{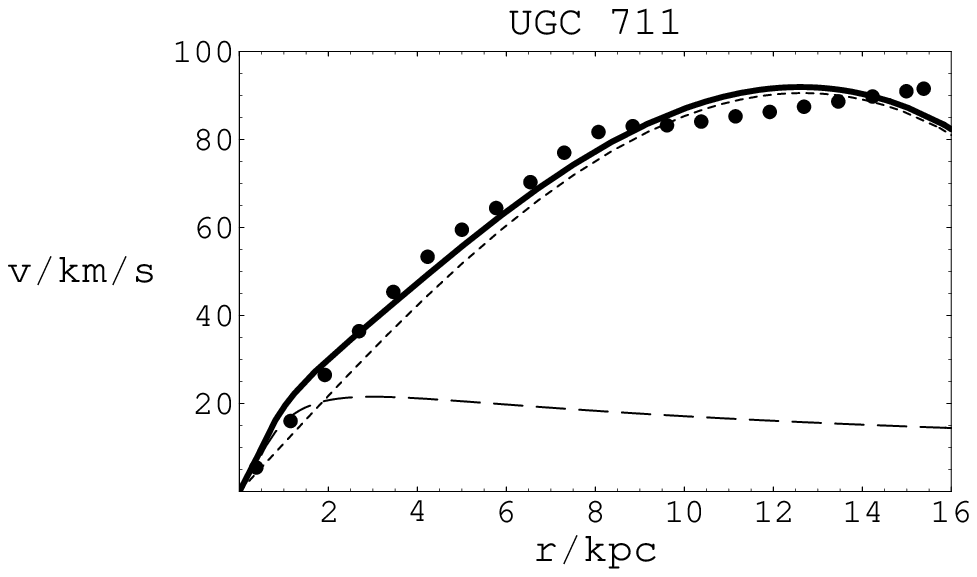}\\[-3cm]
\mbox{}\\[-1.5cm]
\end{figure*}
\begin{figure}
\centering\mbox{}\\[-1cm]
\includegraphics[totalheight=0.35\textheight,width=.48\textwidth]{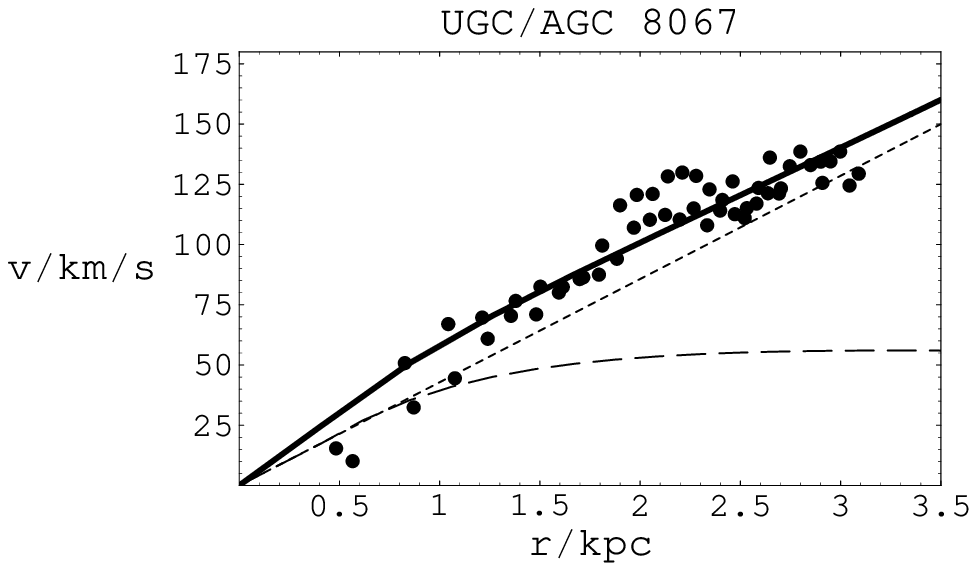}
\includegraphics[totalheight=0.35\textheight,width=.48\textwidth]{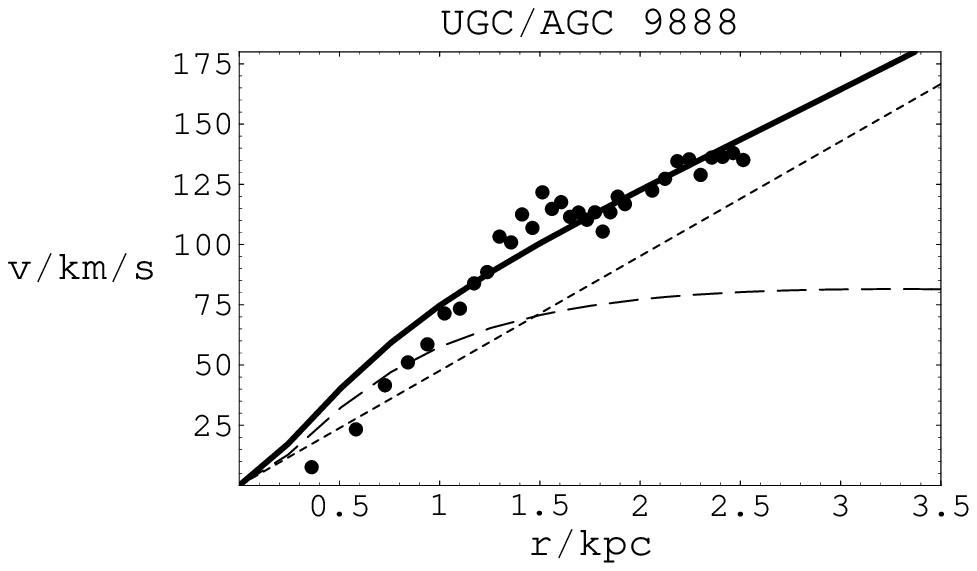}\\[-3cm]
\mbox{}\\[1cm]\caption{Parametric fits of a sample of 12 galactic
rotation curves from the data presented in \cite{deBl02} and
\cite{Spek05}. The points represent the observational values. The
solid curve is the rotation curve determined with the use of Eq.
(\ref{eqexp}). The dotted curve represent the tangential velocity
of the Bose-Einstein condensate, while the dashed curve is the
Newtonian galactic rotation curve, corresponding to the presence
of the normal baryonic matter in the galactic core.}
\label{ddo189-ngc4395}
\end{figure}

\begin{table}[!ht]
\begin{tabular}[t]{|c|c|c|c|c|} \hline \hline
Galaxy & $\quad M_B^{tot}/10^{10}M_{\odot}\quad$ & $\quad R/{\rm
kpc}\quad$ & $\quad \rho _{DM}^{(c)}/(10^{-25}{\rm g/cm^3})\quad$
&
$\quad\chi^2\quad$ \\
\hline
DDO 189 & 0.39 & 7.88 & 5.51 & 0.33 \\
\hline
NGC 100 & 0.44 & 8.38 & 10.39 & 1.64 \\
\hline
NGC 1560 & 0.35 & 9.86 & 5.82 & 5.42 \\
\hline
NGC 4395 & 0.71 & 10.22 & 4.81 & 11.07 \\
\hline
NGC 5023 & 0.70 & 5.67 & 15.13 & 2.31 \\
\hline
UGC 711 & 0.093 & 14.49 & 4.56 & 2.76 \\
\hline
UGC 3371 & 0.17 & 11.22 & 6.12 & 1.92 \\
\hline
UGC 5005 & 0.28 & 24.65 & 2.056 & 4.24 \\
\hline
UGC/AGC 8067$^{\ast}$ & 0.70 & 686.83 & 69.25 & 58.32 \\
\hline
UGC/AGC 9888$^{\ast}$ & 1.49 & 557.39 & 85.49 & 36.014 \\
\hline
UGC 10310 & 0.086 & 8.47 & 9.03 & 0.77\\
\hline
UGC/AGC 241294$^{\ast}$ & 0.94 & 950.31 & 54.81 & 61.015 \\
\hline \hline
\end{tabular}
\caption{Numerical values of the fitting parameters for the fitted
rotation curves. Note that $\chi^2$ is relatively small which suggests
that our model fits the data quite well. In view of the values of $\chi^2$
of the low luminosity dwarf galaxies, marked by $\ast$, we find that the
model fits the high luminosity galaxies much better than the low luminosity
ones. This is rather similar to the standard dark matter models, which also
work best for high luminosity galaxies.} \label{numtable}
\end{table}

As one can see from the figure and from Table~\ref{numtable}, for
$9$ over $12$ cases there is an overall very good agreement
between the data and the best fit curve. This strongly suggest
that our model may be relevant for obtaining a correct description
of the dark matter and its properties. A purely Newtonian
description of these observational data by assuming that the
entire matter is in the form of the baryonic matter concentrated
in a core region is impossible. The Newtonian potential of the
core is asymptotically decreasing, but the corrected rotation
curve is much higher than the Newtonian one, thus offering the
possibility to fit the rotation curves. On the other hand since
LSB galaxies are usually considered to be dark matter dominated,
reproducing their rotation curves represents a significant
evidence in favor of the model. Moreover, fitting to rotation
curves allows to determine the theory parameters, and especially
$R$, the radius of the dark matter halo, and the central density
of the dark matter, $\rho _{DM}^{(c)}$. The values obtained from
the fitting are physically reasonable, indicating a radius of the
galactic halos in the range of $8-25$ kpc. On the other hand, the
values of the central density of the dark matter (at the galactic
center) predicted by our model are of the order of $10^{-25}$
g/cm$^3$, a value much smaller than the value of $\rho
_{DM}\left(100\;{\rm pc}\right)\approx 10M_{\odot }/{\rm
pc}^3\approx 7\times 10^{-22}{\rm g/cm}$$^3$ estimated for our
galaxy \cite{Gn04}.

If our model can fit quite well the observational results for the
LSB galaxies, the fits for the dwarf galaxies based on the data
from \cite{Spek05} require an unrealistically large extension of
the dark matter, and they are statistically not significant. A
possible explanation may be related to the peculiar behavior of
the observational data near the origin: the tangential velocities
do not tend to zero for $r\rightarrow 0$, but are zero for a value
of $r$ of the order of $0.5$ kpc, $v_{tg}\left(0.5\;{\rm
kpc}\right)\approx 0$. Thus, a possibility for a better fitting of
the observational data for these galaxies would be to re-scale the
radial coordinate by introducing a new parameter $r_0$ so that
$r\rightarrow r-r_0$ and $v_{tg}\left(r_0\right)=0$. However, the
necessity of introducing a new parameter seems not to be
physically motivated.

Conventionally, galactic dark matter is modelled by an isothermal
sphere in hydrostatic equilibrium, having two free parameters.
These parameters are adjusted for every galaxy individually. One
severe problem in that approach is that it seems that the
rotational velocity of the last data point scales like the
luminosity. Also, the Tully-Fisher relation $L\sim v_{out}^a$,
$a\approx 4$, where $L$ is the luminosity (in units of
$10^{10}L_{\odot}$) and $v_{out}$ is the velocity at the outermost
observed radial position, which holds for the high luminosity
spiral galaxies \cite{Bi87}, actually suggests that the luminous
matter and the related Newtonian effects must play an important
role in determining the rotational velocities.

\section{Light deflection and lensing by Bose-Einstein condensate dark halos}

One of the ways we could in principle test the galactic dark
matter model obtained in the previous Section would be by studying
the light deflection by galaxies, and in particular by studying
the deflection of photons passing through the region where the
rotation curves are flat. Let us consider a photon approaching a
galaxy from far distances. We will compute the deflection by
assuming that the metric is given by Eq. (\ref{line}), together
with Eqs. (\ref{s4}) -- (\ref{s6}).

The bending of light by the galactic gravitational field results
in a deflection angle $\Delta \phi $ given by
\begin{align}\label{defl1}
\left(\Delta \phi\right)_{BE} =2\left| \phi \left( r_0\right)
-\phi _{\infty }\right| -\pi ,
\end{align}
where $\phi _{\infty }$ is the incident direction and $r_0$ is the
coordinate radius of the closest approach to the center of the
galaxy. Generally \cite{BhKa03}
\begin{align}
\phi \left( r_{0}\right) -\phi _{\infty }=\int_{r_{0}}^{\infty }e^{\frac{%
\lambda (r)}{2}}\left[ e^{\nu \left( r_{0}\right) -\nu \left(
r\right) }\left( \frac{r}{r_{0}}\right) ^{2}-1\right]
^{-1/2}\frac{dr}{r}.
\end{align}

In the dimensionless variables introduced in Eqs. (\ref{var}) we
have
\begin{align}
\phi \left( r_{0}\right) -\phi _{\infty }=\int_{\eta _{0}}^{\infty }\left[1-%
\frac{2M_{0}\left( \eta \right) }{\eta }\right]^{-1/2}\left[
e^{\nu \left( \eta _{0}\right) -\nu \left( \eta \right) }\left(
\frac{\eta }{\eta _{0}}\right) ^{2}-1\right]^{-1/2} \frac{d\eta
}{\eta }.
\end{align}

We shall compare the values of the deflection angle $\Delta $ with
a semi-realistic model for dark matter, in which it is assumed
that the galaxy (the baryonic matter) is embedded into an
isothermal mass distribution (the dark matter),
with the density varying as $\rho =\sigma _{v}^{2}/2\pi Gr^{2}$, where $%
\sigma _{v}$ is the line of sight velocity dispersion \cite{Bi87}.
In this model it is assumed that the mass distribution of the dark
matter is spherically symmetric. In fact, if the rotation curve is
flat, then the mass distribution must be that of the isothermal sphere, for which we also have $%
v_{tg}=\sqrt{2}\sigma _{v}$. The surface density $\Sigma $ of the
isothermal sphere is $\Sigma \left( r\right) =\sigma
_{v}^{2}/2Gr$. For this dark
matter distribution the bending angle of light is constant and is given by $%
\left( \Delta \phi \right) _{DM}=2\pi v_{tg}^{2}/c^2$
\cite{BlNa92}.

To compare the values of the deflection angle in the Bose-Einstein
condensate and isothermal sphere dark matter models we introduce a
dimensionless parameter $\Delta $ defined as
\begin{align}\label{delta}
\Delta =\frac{\left( \Delta \phi \right) _{BE}}{\left( \Delta \phi
\right) _{DM}}=\frac{\left( \Delta \phi \right) _{BE}}{2\pi \left(
v_{tg}/c\right) ^{2}}.
\end{align}

\begin{figure}[!ht]
\includegraphics{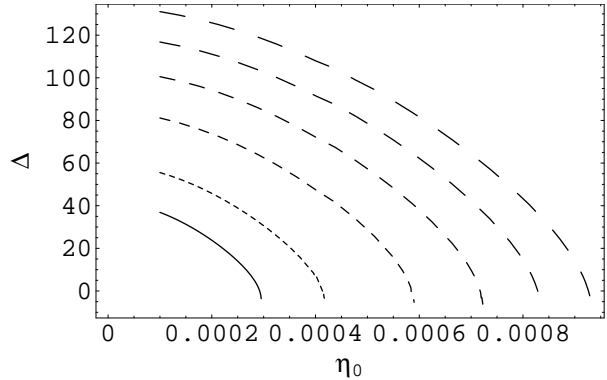}
\caption{The parameter $\Delta $ as a function of the impact
parameter $\eta _0=r_0/r^{\ast}$ for a Bose-Einstein condensate
with $m=5.6\times 10^{-34}$ g and $a=10^6$ fm, for different
values of the central density: $\rho _{DM}^{(c)}=10^{-27}$
g/cm$^3$ (solid curve), $\rho _{DM}^{(c)}=2\times 10^{-27}$
g/cm$^3$ (dotted curve), $\rho _{DM}^{(c)}=4\times 10^{-27}$
g/cm$^3$ (short dashed curve), $\rho _{DM}^{(c)}=6\times 10^{-27}$
g/cm$^3$ (dashed curve), $\rho _{DM}^{(c)}=8\times 10^{-27}$
g/cm$^3$ (long dashed curve) and $\rho _{DM}^{(c)}= 10^{-26}$
g/cm$^3$ (ultra-long dashed curve). In each case $v_{tg}=300$
km/s.} \label{FIG7}
\end{figure}

The variation of $\Delta $ as a function of the impact parameter
$r_0$ is represented, for different values of the central density
of the Bose-Einstein condensate dark matter, in Fig.~\ref{FIG7}.

\begin{figure}[ht]
\includegraphics[width=250pt,height=250pt]{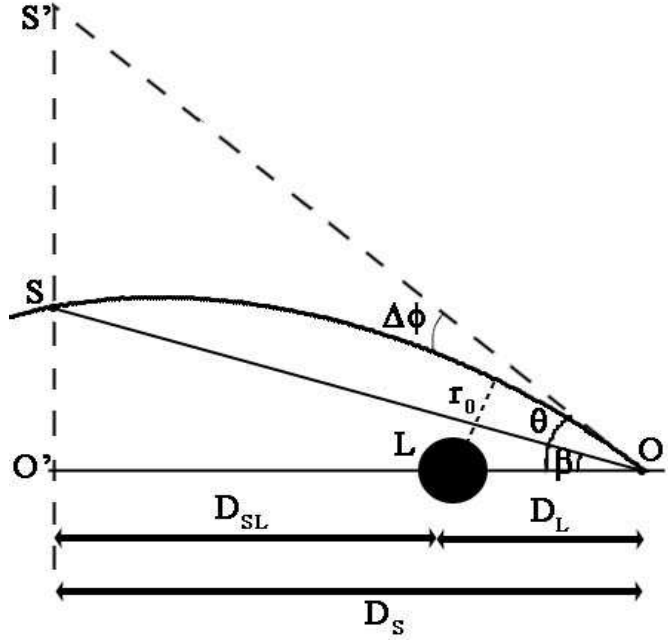}
\caption{The lensing geometry, showing the location of the
observer $O$, of the lensing galaxy $L$ and of the source $S$. The
deflection angle is $\Delta \phi$. The angular diameter distances
$D_L$, $D_{LS}$ and $D_S$ are also indicated.} \label{FIG8}
\end{figure}

Once the light deflection angle is known, one can study the
gravitational lensing by the dark matter halos. The lensing
geometry is illustrated in Fig.~\ref{FIG8}.

The light emitted by the source $S$ is deflected by the lens $L$
(a galaxy in our case) and reaches the observer $O$ at an angle
$\theta $ to the optic axis $OL$, instead of $\beta $. The lens
$L$ is located at a distance $D_L$ to the observer and a distance
$D_{LS}$ to the source, respectively, while the observer-source
distance is $D_S$. $r_0$ is the impact factor (distance of closest
approach) of the photon beam.

The lens equation is given by \citep{BlNa92}
\begin{align}
\tan \beta =\tan \theta -\frac{D_{LS}}{D_{S}}\left[ \tan \theta
+\tan \left( \left(\Delta \phi \right)_{BE} -\theta \right)
\right].
\end{align}

By assuming that the angle $\theta $ is small, we have $\tan
\theta \approx \theta $ and the lens equation can be written as
\begin{align}
\beta \approx \theta -\frac{D_{LS}}{D_{S}}\left(\Delta \phi
\right)_{BE}.
\end{align}

In the special case of the perfect alignment of the source, lens
and observers, $\beta =0$, and the azimuthal axial symmetry of the
problem yields a ring image, the Einstein ring, with angular
radius
\begin{align}
\theta _{E}^{(BE)}\approx \frac{D_{LS}}{D_{S}}\left(\Delta \phi
\right)_{BE}.
\end{align}

This equation can be expressed in a more familiar form by taking
into account that the impact parameter $r_{0}\approx D_{L}\theta
$, which gives
\begin{align}
\theta _{E}^{(BE)}\approx \sqrt{\frac{D_{LS}}{D_{S}D_{L}}\Delta
\phi r_{0}}\approx \theta
_E^{(GR)}\sqrt{\frac{r_0}{GM_B}}\sqrt{\Delta \phi },
\end{align}
where $\theta _E^{(GR)}$ is the angular radius of the Einstein
ring in the case of standard general relativity, $\theta
_{E}^{(GR)}=\sqrt{4\left( D_{LS}/D_{S}D_{L}\right) GM_{B}}$, with
$M_B$ denoting the baryonic mass of the galaxy.

In the case of a galaxy with a heavy isothermal dark matter
distribution, the Einstein radius of the lens formed in perfect
alignment is \citep{BlNa92}
\begin{align}
\theta _{E}^{(DM)}=\left( \frac{4\pi \sigma _{v}^{2}}{c^{2}}\right) \frac{%
D_{LS}}{D_{S}}.
\end{align}

The ratio $\Delta $ of the Einstein's rings angular diameters in
the brane world models and in the isothermal dark galactic halo
model is
\begin{align}
\Delta =\frac{\theta _{E}^{(BE)}}{\theta _{E}^{(DM)}}=\frac{\left(
\Delta \phi \right) _{BE}}{2\pi \left(v_{tg}/c\right)^{2}}.
\end{align}

Therefore the ratio of the angular radii of the Einstein rings in
the Bose-Einstein condensate dark matter model and in the
isothermal dark matter model is given by the same parameter
$\Delta $ which has been already introduced in Eq. (\ref{delta}).
Hence the variation of the ratio of the Einstein rings in the two
models can also be obtained from Fig.~\ref{FIG7}.

\section{Discussions and final remarks}

Galactic rotation curves pose a challenge to present day physics and one
would like to have a better understanding of some of their intriguing phenomena,
like their universality and the very good correlation between the amount of
dark matter and the luminous matter in the galaxy. In the present work
we have considered, and further developed, an alternative view to the
dark matter problem, namely, that the galactic rotation curves can be
explained by models in which dark matter is in the form of a Bose-Einstein
condensate.
This assumption leads to a complete description of the basic properties
of the dark matter condensates at both the Newtonian and general
relativistic level, and gives a definite set of predictions which
can be tested observationally. In particular, we have shown that
the model provides a good descriptions for the galactic rotation
curves by fitting our model to the data of 12 observed curves.

The numerical values of the basic parameters (mass and radius) of
the condensed dark matter halos sensitively depend on the mass $m$
of the condensate, on the scattering length $a$ and on the central
density $\rho _{DM}^{(c)}$, $R_{DM}=R(a,m,\rho _{DM}^{(c)})$,
$M_{DM}=M_{DM}(a,m,\rho _{DM}^{(c)})$. Of course, in
general, the values of the mass and radius of the gravitational
condensate depend on the adopted model for the non-linearity.

The scattering length $a$ is defined as the zero-energy limit of
the scattering amplitude $f$~\citep{Da99}. Depending on the spin
dependence of the underlying particle interaction, the scattering
length may in general be also spin dependent. The spin independent
part of the quantity is referred to as the coherent scattering
length $a_c$. The scattering lengths can be obtained for some
systems in the laboratory. In our estimations we have used the
value of the scattering length obtained for dilute atomic
Bose-Einstein condensates in terrestrial laboratory experiments.
Another essential parameter is the mass $m$ of the condensate
particle, which, due to the lack of information about the physical
nature of the dark matter, is a free parameter, which must be
constrained by observations.

A powerful observational tool for discriminating between standard
dark matter and Bose-Einstein condensate models is the study of the
deflection of light (gravitational lensing) by the dark matter halos.
Due to the fixed form of the galactic metric in
the flat rotation curves region, in standard dark matter models
the light bending angle is a function of the tangential velocity
of particles in stable circular orbit and of the baryonic mass and
radius of the galaxy. Generally, the specific form of the bending
angle is determined by the galactic metric, and this form is very
different for the Bose-Einstein condensate dark matter as compared
to the other dark matter or modified gravity models (MOND
\cite{Mi}, non-symmetric gravity \cite{Mo96}, long-range
self-interacting scalar fields \cite{Ma03} etc.).

The gravitational light deflection angle predicted by the
Bose-Einstein condensate models is much larger than the value
predicted by the standard dark matter approach. For example, there
are significant differences in the lensing effect with respect to
the isothermal dark matter halo model. Therefore the study of the
gravitational lensing may provide evidence for the existence of
the dark matter in the form of a Bose-Einstein condensate.
Therefore, the study of the galaxy-galaxy lensing and of the dark
matter halos' properties could provide strong constraints on the
Bose-Einstein dark matter model and on related physical models.

In the present approach to dark matter all the relevant physical
quantities can be obtained from observable parameters (the dark
halo mass, the radius of the galaxy and the observed flat
rotational velocity curves). Therefore this opens the possibility
of testing the Bose-Einstein condensation models by using
astronomical and astrophysical observations at the
galactic-intergalactic scale. In this paper we have provided some
basic theoretical tools necessary for the in depth comparison of
the predictions of the condensate model and of the observational
results.

\acknowledgments
The work of CGB was supported by research grant BO 2530/1-1 of the
German Research Foundation (DFG). TH is supported by the RGC grant
No.~7027/06P of the government of the Hong Kong SAR.

\appendix
\section{Motion of test particles}
\label{appI}

The motion of a test particle with four-velocity $u^{\mu }$ in the
gravitational field of the galaxy can be described by the
Lagrangian \citep{Ha}
\begin{align}
2L=\left( \frac{ds}{d\tau }\right) ^{2}=-e^{\nu \left( r\right)
}\left(
\frac{dt}{d\tau }\right) ^{2}+e^{\lambda \left( r\right) }\left( \frac{dr}{%
d\tau }\right) ^{2}+r^{2}\left( \frac{d\Omega }{d\tau }\right)
^{2},
\end{align}
where $d\Omega ^{2}=d\theta ^{2}+\sin ^{2}\theta d\phi ^{2}$ and
$d\tau =cdt$. From the Lagrange equations it follows that we have
two constants of motion, the energy $E=e^{\nu (r)}\dot{t}$ and the
angular momentum $l=r^{2}\dot{\phi}$ \citep{La03}.
The condition $u^{\mu }u_{\mu }=-1$ gives $-1=-e^{\nu \left( r\right) }\dot{t%
}^{2}+e^{\lambda (r)}\dot{r}^{2}+r^{2}\dot{\phi}^{2}$ and, with
the use of the constants of motion we obtain
\begin{align}\label{energy}
E^{2}=e^{\nu +\lambda }\dot{r}^{2}+e^{\nu }\left( \frac{l^{2}}{r^{2}}%
+1\right).
\end{align}

This equation shows that the radial motion of the particles on a
geodesic is the same as that of a particle with position dependent mass and with energy $%
E^{2}/2$ in ordinary Newtonian mechanics moving in the effective potential $%
V_{eff}\left( r\right) =e^{\nu (r)}\left( l^{2}/r^{2}+1\right) $.
The conditions for circular orbits $\partial V_{eff}/\partial r=0$
and $\dot{r}=0 $ lead to \citep{La03}
\begin{align}\label{cons}
l^{2}=\frac{1}{2}\frac{r^{3}\nu ^{\prime }}{1-\frac{r\nu ^{\prime }}{2}}%
,E^{2}=\frac{e^{\nu }}{1-\frac{r\nu ^{\prime }}{2}}.
\end{align}

The line element given by Eq. (\ref{line}) can be rewritten in
terms of the spatial components of the velocity, normalized with
the speed of light, measured by an inertial observer far from the
source, as $ds^{2}=-c^2dt^{2}\left( 1-v^{2}/c^2\right) $
\cite{Ha}, where
\begin{align}
\frac{v^{2}}{c^2}=e^{-\nu }\left[ e^{\lambda }\left(
\frac{dr}{cdt}\right) ^{2}+r^{2}\left( \frac{d\Omega
}{cdt}\right)^{2}\right].
\end{align}

For a stable circular orbit $\dot{r}=0$, and the tangential
velocity of the test particle can be expressed as
\begin{align}
\frac{v_{tg}^{2}}{c^2}=\frac{r^{2}}{e^{\nu }}\left( \frac{d\Omega
}{cdt}\right)^{2}.
\end{align}

In terms of the conserved quantities the angular velocity is given
by
\begin{align}
\frac{v_{tg}^{2}}{c^2}=\frac{e^{\nu }}{r^{2}}\frac{l^{2}}{E^{2}}.
\end{align}

With the use of Eqs. (\ref{cons}) we obtain
\begin{align}
\frac{v_{tg}^{2}}{c^2}=\frac{r\nu ^{\prime }}{2}.
\end{align}

\end{document}